\documentclass[journal=jacsat,manuscript=article]{achemso}

\usepackage[version=3]{mhchem} 

\usepackage{amsmath}
\usepackage{amssymb}
\usepackage{color}
\usepackage{CJK}
\usepackage{framed}
\usepackage{hyperref}
\usepackage{algorithm}
\usepackage{algpseudocode}
\algdef{SE}[DOWHILE]{Do}{doWhile}{\algorithmicdo}[1]{\algorithmicwhile\ #1}


\newcommand*{\blauw}[1]{#1}
\newcommand*{\groen}[1]{#1}

\newcommand*{\noter}[1]{}					

\author{Jiang Wang}
\email{jwang110@illinois.edu}
\affiliation{Department of Physics, University of Illinois Urbana-Champaign, Urbana, IL 61801, USA}
\phone{(217) 244-8487}
\fax{(217) 333-2736}

\author{Andrew L. Ferguson}
\affiliation{Department of Materials Science and Engineering, University of Illinois Urbana-Champaign, Urbana, IL 61801, USA}
\alsoaffiliation{Department of Chemical and Biomolecular Engineering, University of Illinois Urbana-Champaign, Urbana, IL 61801, USA}
\alsoaffiliation{Department of Physics, University of Illinois Urbana-Champaign, Urbana, IL 61801, USA}
\phone{(217) 244-8487}
\fax{(217) 333-2736}

\title[Simulation and machine learning of ring polymers]{A study of the morphology, dynamics, and folding pathways of ring polymers with supramolecular topological constraints using molecular simulation and nonlinear manifold learning}


\begin{document}

\newpage

\begin{abstract}

\noindent  Ring polymers are prevalent in natural and engineered systems, including circular bacterial DNA, crown ethers for cation chelation, and mechanical nanoswitches. The morphology and dynamics of ring polymers are governed by the chemistry and degree of polymerization of the ring, and intramolecular and supramolecular topological constraints such as knots or mechanically-interlocked rings. In this study, we perform molecular dynamics simulations of polyethylene ring polymers at two different degrees of polymerization and in different topological states, including a trefoil knot, catenane state (two interlocked rings), and Borromean state (three interlocked rings). We employ nonlinear manifold learning to extract the low-dimensional free energy surface to which the structure and dynamics of the polymer chain are effectively restrained. The free energy surfaces reveal how degree of polymerization and topological constraints affect the thermally accessible conformations, chiral symmetry breaking, and folding and collapse pathways of the rings, and present a means to rationally engineer ring size and topology to stabilize particular conformational states and folding pathways. We compute the rotational diffusion of the ring in these various states as a crucial property required for the design of engineered devices containing ring polymer components. 
\end{abstract}


\newpage

\section{Introduction}

Ring polymers exist widely in nature, including circular bacterial DNA \cite{Vologodskii1963, Helinski1971, Fain1997} and crown ethers \cite{Crown}. Synthetic ring polymers are useful components in the engineering of artificial micro-devices and molecular machines, such as catenane nanoswitches made up of two interlocked ring polymers \cite{Sauvage2007, balzani2000artificial, amabilino1995interlocked, collier20002, balzani1998molecular, pease2001switching, raymo1999interlocked}. Due to the absence of free ends, the topology of ring polymers gives rise to significant structural and dynamic differences compared to their linear counterparts, and the structure and rheological properties of ring polymers have been subject to extensive theoretical \cite{Rubinstein1986, theory2, theory3, hsiao2016ring}, experimental \cite{experiment1, experiment2, experiment3}, and simulation \cite{simulation1, simulation2, simulation3, simulation4} investigations. 

For linear polymers in the melt state, there is a transition to entanglement when the chain length is greater than a critical value $N_C$ \cite{Gennes1971, Rubinstein2003, McLeish2002}. The relaxation of linear entangled polymers is well described by the reptation model, in which the free end of the linear chain reptates (slides) along the tube-like walls formed by other entangled polymers. For ring polymers in which there are no free ends, alternative models and investigations are needed to describe the polymer dynamics. A theoretical investigation of ring polymers conducted by Rubinstein showed that the dynamics of non-concatenated polymer rings inside crosslinked polymer gels is much faster than that of entangled branched polymers, and is similar to the dynamics of entangled linear polymers \cite{Rubinstein1986}. Experimental work by Pasquino \textit{et al.} measured the linear rheology of critically purified ring polyisoprenes, polystyrenes, and polyethyleneoxides of different molar masses, to reveal that rings exhibit a universal trend clearly departing from that of their linear counterparts \cite{experiment1}. A number of computational studies have also been performed. Brown and Szamel employed the bond fluctuation model to show that -- due to the lack of one degree of freedom -- the average size of the ring polymer is smaller than a linear chain with the same degrees of polymerization, such that the ring is more compact, has larger fractal dimension, and possesses an elevated diffusivity relative to linear chains in the melt \cite{simulation3, simulation4}. Yang \textit{et al.}\ studied the diffusion of 30-ring polymer in a matrix of linear chains using Monte Carlo simulation, where they find that the threading of a linear polymer into the ring would significantly slows the diffusion of the ring \cite{simulation2}. Recent molecular dynamics simulation study by Jung and co-workers also show that the threading between the rings can underlie the slowing down in diffusivity as ring size increases \cite{Jung2015}.



Compared to linear chains, the literature on the properties of isolated ring polymers is relatively sparse. For single polymers in solvent with no entanglement, Schroeder and co-workers studied the dynamics of DNA chains in hydrodynamic flows \cite{hsiao2016direct, Zhou2016}. Sing and co-workers combined Brownian dynamics simulations, analytical theory, and experiment to study ring polymers in flow, and investigate a transition from the large loop conformation at intermediate Weissenberg numbers to the coil-stretch transition at larger Weissenberg numbers \cite{hsiao2016ring}. Chen \textit{et al.} studied the behavior of single flexible ring polymers in simple shear flow using multiparticle collision dynamics combined with molecular dynamics simulation to study the mechanisms of the two fundamental motions of ring polymers: tumbling and tank-treading \cite{chen2015conformations}.


The closed-loop structure of ring polymers makes it possible to construct supramolecular assemblies of interlocked loops, wherein the constituent ring polymers are mutually connected by mechanical bonds \cite{Sauvage2007}. The 2016 Nobel Prize in Chemistry recognized Stoddart, Sauvage, and Feringa for pioneering work in the synthesis of such supramolecular chemistries as the basis for molecular machines \cite{2016nobel}. Examples of experimentally synthesized supramolecular topologies include knots, catenane, Borromean ring, and rotaxane assemblies \cite{Sauvage2007, balzani2000artificial, amabilino1995interlocked, collier20002, balzani1998molecular, pease2001switching, raymo1999interlocked, Link2016}. The additional constraints introduced by the supramolecular topology perturbs the structure and dynamics of the constituent rings. For example, it is a mathematical truth that the rings constituting a Borromean ring cannot be perfectly circular \cite{freedman1987, Linstrom1991}. A small number of recent studies have explored the dynamics and configurations of concatenated ring polymers. Polles \textit{et al.} conducted simulations of the self-assembly of rigid links in slit pores, and found that links assembled from helical templates cover a rich, but very specific, repertoire of topologies \cite{Polles2016}. Fernando \textit{et al.} synthesized [2]catenanes with electron-rich rings, and performed quantum mechanical calculations to show that the oxidation state can be used to control the energy barriers for the relative motion of the interlocked rings \cite{slidecatenane}. Despite these advances, there is a pressing need for more systematic and generalized studies of the configurational behavior and dynamics of ring polymers in various supramolecular topologies.


In this work, we quantify the influence of topological constraints on the structural and dynamic behavior of ring polymers within supramolecular assemblies using molecular dynamics simulations and nonlinear dimensionality reduction. By resolving the position of each constituent atom in the system as a function of time, molecular dynamics simulation provides a means to determine the conformation and dynamics of the ring polymers at high time and spatial resolution, and incorporate the multi-body molecular interactions governing the details of system behavior through validated molecular force fields \cite{Frenke2001un}. Nonlinear dimensionality reduction techniques provide a means to extract from high-dimensional molecular simulation trajectories the important collective variables governing the long-time dynamical evolution of the system \cite{ferguson2011cpl}. Low-dimensional free energy surfaces constructed in these collective variables can resolve the important metastable states of the molecular system and the low-free energy folding pathways by which they are connected, providing deep insight into the molecular thermodynamics and mechanisms. 

We adopt as our system of study 24-mer and 50-mer polyethylene chains in water as prototypical models of a hydrophobic polymer \cite{ferguson2010systematic, Rachael_2015}, and study their behavior in a variety of supramolecular topologies including linear chains, rings, trefoil knots, catenanes, and Borromean rings. Despite its chemical simplicity, the structure and dynamics of polyethylene chains is very rich, and it offers a clean test system with which to resolve the impact of supramolecular topology upon chain behavior without the confounding effects of more complicated monomer chemistries.

The remainder of this paper is organized as follows. In \blauw{Section 2}, we provide details of our molecular simulation approach and nonlinear dimensionality reduction technique. In \blauw{Section 3}, we report the low-dimensional molecular free energy landscapes for the two chain lengths in each supramolecular assembly, and report our analysis of chain structure, energetics, relaxation rates, and rotational diffusivity. In \blauw{Section 4}, we present our conclusions and outlook for future work.

\section*{2. Methods}

\subsection*{2.1 Molecular dynamics simulation}

\begin{figure*}[ht!]
\includegraphics[width=0.99\textwidth]{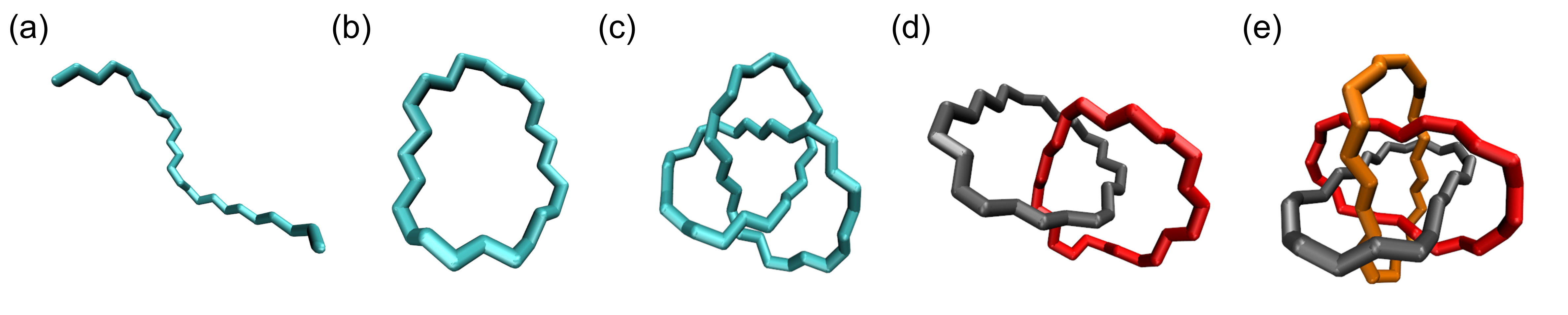}
\caption{\label{fig1} Schematic illustration of the six (supramolecular) topologies considered in this work illustrated for the polyethylene molecule: (a) 24-mer linear chain, (b) 24-mer ring, (c) 50-mer trefoil knot in left and right chiralities (left chirality shown here), (d) 24-mer catenane, (e) 24-mer Borromean ring.}
\end{figure*}

Molecular dynamics simulations were conducted using GROMACS 4.6 simulation suite \cite{Gromacs}. We studied polyethylene chains at two different degrees of polymerization, the smaller containing 24 carbon atoms and the larger 50 carbon atoms. For each chain length, we considered six different supramolecular topologies as illustrated in \blauw{Fig.~\ref{fig1}}: (a) linear chain, (b) ring, (c) trefoil knot in left and right chiralities, (d) catenane comprising two identical interlocked rings, and (e) Borromean ring comprising three interlocked rings. We recall that the Borromean ring fundamentally differs from the catenane in that no pair of rings are interlocked, but the triplet of rings is mechanically linked \cite{Sauvage2007}. We constructed the polymer topologies with the assistance of the PRODRG2 server \cite{PRODRG}, and modeled their interactions using the united atom TraPPE potential \cite{Martin1998}. Each topology was placed in a cubic box and solvated by SPC water molecules to a density of 1.0 g/cm$^3$ \cite{water}. Periodic boundary conditions were implemented in all three dimensions, and box side lengths between 3-5 nm depending on the system to preclude direct interactions between the polymer chains through the periodic walls. Lennard-Jones interactions were smoothly shifted to zero at 1.4 nm, and Lorentz-Berthelot combining rules applied to determine interactions between unlike atoms \cite{allen1989computer}. Electrostatics were treated by particle mesh Ewald with a real-space cutoff of 1.4 nm and reciprocal-space grid spacing of 0.12 nm \cite{essmann1995smooth}. All bond lengths were maintained at their equilibrium length using the LINCS algorithm \cite{Hess1997}. High energy overlaps in initial system configurations were relaxed by steepest descent energy minimization to eliminate forces exceeding 2000 kJ/mol.nm. Simulations were conducted in the NPT ensemble with temperature maintained at 298 K using a Nos\'{e}-Hoover thermostat \cite{nose1984unified} and 1 bar using an isotropic Parinello-Rahman barostat \cite{parrinello1981polymorphic}. The classical equations of motion were numerically integrated using the leap-frog algorithm \cite{hockney2010computer} with a 2 fs time step. Each system was equilibrated for 1 ns, at which time temperature, pressure, energy, and structural averages had all stabilized. Production runs of 100 ns were then conducted, and system snapshots harvested every 1 ps for offline analysis using in-house codes developed in MATLAB R2013b (The MathWorks Inc., Natick, MA) and C++. Simulation trajectories were visualized using VMD \cite{humphrey1996vmd}.

\subsection*{2.2 Diffusion maps manifold learning}

The molecular dynamics simulation trajectories describe a realization of the time evolution of the system within the 3$N$-dimensional configurational phase space comprising the three-dimensional Cartesian coordinates of the $N$ atoms in the system. Due to interactions between the atomic degrees of freedom mediated by intra and intermolecular forces, the effective dimensionality of a molecular system is far lower than the 3$N$-dimensional space in which the equations of motion are formulated \cite{ferguson2011cpl, ferguson2010systematic, Ferguson_takens, Angel1992, amadei1993, hegger2007, zhuravlev2009, das2006}. Geometrically, this can be understood as the emergence of a small number of collective variables governing the long-time evolution of the system to which the remaining degrees of freedom are slaved \cite{ferguson2011cpl, ferguson2010systematic, zwanzig2001, hummer2003, cho2006}. We and others have previously demonstrated that these collective variables can be determined from molecular simulation trajectories using nonlinear manifold learning \cite{nadler2006advances, best2010coordinate, lafon2006data, singer2009detecting, rohrdanz2011determination, coifman2006diffusion, coifman2008diffusion, lpbeltrami, coifman2005geometric, ferguson2011integrating, zheng2011polymer, Rachael_2015, Ferguson_takens, Long_2014, Long_2015, Long_2016, ferguson2011cpl, ferguson2010systematic}, the particular variant of which we use here are diffusion maps \cite{coifman2006diffusion, coifman2005geometric, coifman2008diffusion, nadler2006advances, lpbeltrami}. In a nutshell, the diffusion map approach constructs a random walk over the high-dimensional simulation trajectory with hopping probabilities based on the structural similarity of the constituent snapshots, then performs a spectral analysis of the harmonics of the resultant discrete Markov process to ascertain the effective dimensionality of the underlying ``intrinsic manifold'' and nonlinear collective variables with which to parameterize it \cite{ferguson2011cpl}. We have previously described the application of diffusion maps to determine low-dimensional representations of the molecular free energy landscape for linear chains \cite{ferguson2010systematic, Rachael_2015, Ferguson_takens}. We briefly summarize the technique below, including a discussion of the specializations required in applications to ring homopolymers.

We harvest from our simulation trajectories the ensemble of $N$ configurations of the polymer chain in all six topologies (\blauw{Fig.~\ref{fig1}}). In the case of systems containing multiple polymer chains (i.e., catenane, Borromean ring), we consider the configurations of one arbitrarily selected chain. The first step in applying diffusion maps is to compute structural distances $d_{i,j}$ between all pairs of chain configurations $i$ and $j$. We note that in neglecting the coordinates of the solvent and other chains in computing these distances, we treat the chain environment as an external perturbation that manifests itself in the ensemble of thermally accessible chain configurations \cite{ferguson2010systematic}. For linear homopolymers $d_{i,j}$ is straightforwardly computed as the rotationally minimized root mean square deviation (\blauw{RMSD}) between the (united) atom coordinates, that is efficiently calculated using the Kabsch algorithm \cite{ferguson2011cpl, ferguson2010systematic, ferguson2010experimental, kabsch1976solution}. The indistinguishability of the constituent monomers in a ring homopolymer requires that we augment the distance definition with a minimization over indexing of the united atoms composing the chain. In practice, we apply the Kabsch algorithm under each possible indexing and select the minimum. We provide a proof in the \blauw{Supplementary Information} that distances defined in this manner satisfy the triangle inequality, and -- combined with their non-negativity and symmetry -- therefore provide a proper metric function over the molecular configurations \cite{harary6graph, khamsi2011introduction_Chap1pt3}.

Having computed all pairwise distances, the diffusion map approach then proceeds by convoluting each distance with a Gaussian kernel $A_{ij}=\exp(-d_{ij}^2/2\epsilon)$, where $\sqrt{\epsilon}$ is the kernel bandwidth that defines the characteristic hop size. We tune $\epsilon$ to lie in the linear region of the log-log plot of $\sum_{i,j}A_{ij}(\epsilon)$, such that $\epsilon$ is neither too small and the graph becomes disconnected, nor too large and the graph becomes fully connected \cite{ferguson2011integrating, ferguson2011cpl, Coifman2008}. We then row normalize the rows of the Guassian convoluted pairwise distances matrix $\mathbf{A}$ to define a right-stochastic Markov matrix $\mathbf{M} = \mathbf{D}^{-1}\mathbf{A}$, where $\mathbf{D}$ is a diagonal matrix with elements $D_{ii} = \sum_k A_{ik}$, and $M_{ij}$ defines the hopping probability from state $i$ to state $j$ under one application of the discrete random walk \cite{nadler2006advances}. Finally, we perform a spectral decomposition of the Markov matrix $\mathbf{M}$ to obtain the ordered right eigenvectors $\{\vec{\psi_i}\}_{i=1}^N$ and eigenvalues $\{\lambda_i\}_{i=1}^N$ in non-ascending order, with $\lambda_i \in (0,1]$ and $\lambda_1 = 1$ and $\vec{\psi_1} = \vec{1}$ the trivial pair corresponding to the steady state distribution of the random walk. The leading eigenvectors are discrete approximations to the top eigenfunctions of a backward Fokker-Planck equation \cite{Coifman2008}, and therefore correspond to the slow collective modes governing the long-time evolution of the random walk over the data. The diffusion map embedding identifies a spectral gap in the eigenvalue spectrum after the $(k+1)^{th}$ eigenvalue, and projects each of the $i=1 \ldots N$ observed chain configurations into the top $k$ non-trivial eigenvectors,
\begin{equation}
\text{configuraton}_i \rightarrow \left( \vec{\psi}_2(i), \vec{\psi}_3(i), \ldots \vec{\psi}_{k+1}(i)  \right) \label{eqn:dMapEmbed}
\end{equation}
This projection defines the so-called ``intrinsic manifold'' to which the system dynamics are effectively restricted and which contains the important dynamical motions governing the long-time dynamical evolution \cite{ferguson2010systematic, ferguson2011cpl, coifman2006diffusion, coifman2005geometric}. By collating the chain configurations from simulations at all six topologies in this calculation, we define a ``composite'' diffusion map intrinsic manifold that provides a unified low-dimensional projection (i.e., \blauw{Eqn.~\ref{eqn:dMapEmbed}}) spanning the configurational ensembles explored under all conditions \cite{Rachael_2015, Long_2014, Long_2015}. By projecting the data recorded in each individual simulation into this embedding, we can compare how the supramolecular environment affects the extent and distribution of chain configurations over the underlying intrinsic manifold within a unified low-dimensional basis.

\subsection*{2.3 Accelerated nonlinear manifold learning using pivot diffusion maps (P-dMaps)}

Application of diffusion maps to $N$ chain configurations requires calculation of $N(N-1)/2$ pairwise distances, which can become computationally prohibitive particularly when collating data collected from multiple simulations. One may instead construct an approximate diffusion map embedding by considering a subset of $n \ll N$ points, and then project in the remaining $(N-n)$ points using an out-of-sample extension technique such as the Nystr\"{o}m extension \cite{nystromPRE, nystrom1, fowlkes2004spectral, lafon2006data, Baker1977_Numerical}. Significant savings in computation and memory are realized since one now has to store and diagonalize a much smaller $n \times n$ matrix, and the complexity of the out-of-sample extension for each point is $\mathcal{O}(n)$. Accurate approximations to the full $N$-point diffusion map embedding requires judicious selection of the $n$ ``pivot'' points to satisfy two criteria based on the size of the Gaussian kernel bandwidth used in the construction the diffusion map. First, each pivot should be within a distance of $\sqrt{\epsilon}$ from at least one other pivot to ensure that the resulting diffusion map constructed over the $n$ pivots does not become disconnected, thereby frustrating the determination of a single unified embedding. Second, all $(N-n)$ out-of-sample points should lie within $\sqrt{\epsilon}$ of at least one pivot such that their Nystr\"{o}m embedding is accurate \cite{ferguson2011cpl, Bengio2003}. Both of these criteria are satisfied by tiling the high-dimensional point cloud with $n$ pivots contained within hyperspheres of radius $\sqrt{\epsilon}$ or less. In principle, the optimal selection of pivots could be computed from the $N \times N$ pairwise distances matrix, but this would defeat the objective to avoid computation of all distances. The landmark Isomap approach demonstrated good performance of the Isomap nonlinear manifold learning approach by random selection of pivots \cite{Tenenbaum_2000,deSilva2002_NIPS}, but this is expected to fail for datasets containing large density variations since pivots will be less prevalent in low-density regions. A sophistication of this approach employed L1-regularizion of an objective function to perform automated pivot selection \cite{Silva2005_NIPS}, and fast tree search algorithms can efficiently identify nearest neighbors to inform pivot selection \cite{McQueen2016_JMachLearnRes, Muja2014_TPAMI}. Here we propose an alternative approach to on-the-fly pivot selection that is simple to implement, does not require calculation of the full pairwise distances matrix, and its termination automatically furnishes the $n^2$ pairwise distances between pivots required to compute the approximate diffusion map embedding, and the the $n(N-n)$ distances required to perform Nystr\"{o}m embedding of the remaining non-pivot points. We term this approach ``pivot diffusion maps'' (\groen{P-dMaps}).

The idea of the approach is very simple. At any stage during the application of the algorithm, each point is assigned to one of three states: \textbf{0}, \textbf{1}, or \textbf{2}. State \textbf{1} means that the point has been identified as a pivot. State \textbf{2} means that the point is within the domain of an identified pivot, and it is not a pivot. State \textbf{0} means that the point has not yet been classified as \textbf{1} or \textbf{2}. Initially, all points are in state \textbf{0}. We consider each unclassified point in turn and perform the following operations. Upon reaching a new point in state \textbf{0}, we first mark it as a new pivot by switching it to state \textbf{1}, compute pairwise distances between this newly assigned pivot and all $N$ points in the ensemble, and save these distances to the matrix $\mathbf{d}_\mathrm{Nystrom}$. From these distances, we identify all points with state \textbf{0} within a distance $r \leqslant r_\mathrm{cut}$ of this newly assigned pivot, and switch their states to \textbf{2}. We then repeat this analysis for the next state \textbf{0} point in the ensemble. At the termination of this procedure, all points have been assigned to either state \textbf{1} or \textbf{2}. The $n$ points in state \textbf{1} are the pivots, and the distances between these pivots and all $N$ points in the ensemble are stored in the $n$-by-$N$ matrix $\mathbf{d}_\mathrm{Nystrom}$. We then extract from this matrix the $n$-by-$n$ matrix $\mathbf{d}_\mathrm{pivot}$ containing the pairwise distances between pivots. We apply diffusion maps to the $\mathbf{d}_\mathrm{pivot}$ matrix using a kernel bandwidth $\sqrt{\epsilon} \geq r_\mathrm{cut}$ to determine a low-dimensional embedding of the pivots. The remaining $(N-n)$ non-pivot points are then projected into the intrinsic manifold defined by the pivots using the Nystr\"{o}m extension \cite{nystromPRE, nystrom1, fowlkes2004spectral, lafon2006data, Baker1977_Numerical}. Pseudo code for the P-dMaps algorithm is presented in \blauw{Algorithm~\ref{alg:pivot}}. 

We make two observations about the algorithm. First, we note that the outcome of the algorithm (i.e., the assignment of the pivots and resulting low-dimensional embedding) depends on the order in which the unclassified points are considered and assigned. By design, distances between neighboring pivots are $\sim$$r_\mathrm{cut}$ so that all $(N-n)$ points to be projected find themselves at most $r_\mathrm{cut} \leq \sqrt{\epsilon}$ away from a pivot, and can be accurately embedded by the Nystr\"{o}m extension \cite{ferguson2011cpl, Bengio2003}. Provided that the manifold is well covered by pivot points within $r_\mathrm{cut} \leq \sqrt{\epsilon}$ hyperspheres, then the embedding of the $(N-n)$ non-pivot points is robust to the precise selection of pivots. Accordingly, the unclassified points may be considered in any order. Second, we are at liberty to select any $r_\mathrm{cut} \leq \sqrt{\epsilon}$. In practice we select $r_\mathrm{cut} = 0.9 \times \sqrt{\epsilon}$ to efficiently populate the manifold with a small number of pivots while ensuring that they are mutually visible to one another within the bandwidth of the diffusion map kernel. Smaller values of $r_\mathrm{cut}$ unnecessarily increase the number of pivots, and in the limit $r_\mathrm{cut} \rightarrow 0$ we have $n \rightarrow N$ such that every point is a pivot and we recover the original diffusion map formulation.

\begin{figure}[ht!]
\begin{center}
\begin{algorithm}[H]
    \caption{Pivot diffusion maps}
    \label{alg:pivot}
    \begin{algorithmic}[1] 
	\Procedure{P-dMaps}{}
	\State $P(k) \gets 0 \qquad \qquad \; k=1 \ldots N$	\Comment{Initialize all $N$ points to state \textbf{0}}
	\State $n \gets 0$	\Comment{Initialize number of pivots to zero}
	\State $d_\mathrm{Nystrom}(p,q) = []$	\Comment{Initialize empty Nystr\"{o}m distances matrix}
	\For{$i$ := $1$ to $N$} 
		\If{$P(i) = 0$} 
		\State $P(i) \gets 1$ 	\Comment{Assign point $i$ to be a new pivot}
		\State $n \gets n + 1$	\Comment{Increment number of pivots by one}
		\For{$j$ := $1$ to $N$}
			\State $d_\mathrm{Nystrom}(n,j) \gets d_{n,j}$	\Comment{Computing and storing distances to new pivot}
			\If{$((d_\mathrm{Nystrom}(n,j) \leq r_\mathrm{cut})$ and $(P(j) = 0))$}
				\State $P(j) \gets 2$ 	\Comment{Identifying points within $r_\mathrm{cut}$ as non-pivots}
			\EndIf
		\EndFor
		\EndIf		
	\EndFor
	\State $d_\mathrm{pivot} \gets d_\mathrm{Nystrom}(:,P(k) = 1) \qquad \qquad \; k=1 \ldots N$ 	\Comment{Extract $d_\mathrm{pivot}$ from $d_\mathrm{Nystrom}$}
	\State Compute diffusion map embedding of the $n$ pivots using $d_\mathrm{pivot}$
	\State Project the $(N-n)$ non-pivots into the embedding using the Nystr\"{o}m extension
    \EndProcedure
    \end{algorithmic}
\end{algorithm}
\end{center}
\end{figure}

\subsection*{2.4 Free energy surfaces over the intrinsic manifold}

The free energy surfaces (\groen{FES}) supported by the intrinsic manifold for each supramolecular topology can be straightforwardly estimated from the distribution of observations of chain configurations projected into the P-dMaps embedding from that particular molecular simulation \cite{ferguson2010systematic, Rachael_2015}. Specifically, assuming that the simulation is sufficiently long to sample an equilibrium distribution, we estimate the free energy surface parameterized by the P-dMaps eigenvectors using the statistical mechanical expression, $\beta G(\{\vec{\psi}_i\}) = -\log \hat{P}(\{\vec{\psi}_i\}) + C$, where $\beta = 1/k_B T$ is the inverse temperature, $k_B$ is Boltzmann's constant, $T$ is the temperature, and $C$ an arbitrary additive constant reflecting our ignorance of the absolute free energy scale \cite{Rachael_2015, Long_2015, Long_2016}. $G(\{\vec{\psi}_i\})$ is the free energy estimate at the point $\{\vec{\psi}_i\}$, and in the present case is identifiable as the Gibbs free energy since calculations were conducted in the NPT ensemble. $\hat{P}(\{\vec{\psi}_i\})$ is the numerical estimate of the probability density over the manifold. This function is typically estimated by collecting histograms over the points embedded into the low-dimensional projection, but this discrete approach proves unsatisfactory for ramified or rugged free energy landscapes where the resultant histograms can be highly sensitive to bin size and placement. Accordingly, we instead employ a grid-free approximation that naturally conforms to the structure of the landscape by depositing multivariate Gaussians on each point and estimating $\hat{P}(\{\vec{\psi}_i\})$ as the normalized sum of all Gaussians at each point on the manifold \cite{Wang2017}. In this approach, the variance of the Gaussians play an analogous role to bin size by controlling the tradeoff between resolution of details of the landscape and sufficient overlap between Gaussians to provide robust statistics. We have previously shown that standard deviations $\sigma_k \approx L_k/100$, where $L_k$ is the linear extent of the manifold in the dimension spanned by $\vec{\psi}_k$, provides a good rule of thumb that tends to produce satisfactory results.

\section*{3. Results and Discussion} 

\subsection*{3.1 Structure, energy, and free energy surfaces for 24-mer chains}

\textbf{Configurations}. We analyzed using P-dMaps the 60,000 chain configurations of the 24-mer chain harvested from 100 ns molecular dynamics simulations in each of the six topologies illustrated in \blauw{Fig.~\ref{fig1}}. A gap in the eigenvalue spectrum after eigenvalue $\lambda_5$ informed the construction of 4D nonlinear embeddings in the top four non-trivial eigenvectors $( \vec{\psi}_2, \vec{\psi}_3, \vec{\psi}_4, \vec{\psi}_5 )$ (\blauw{Eqn.~\ref{eqn:dMapEmbed}}, \blauw{Fig.~S1a}). Consistent with our previous work, we find that the principal moments of the gyration tensor $\{\xi_1,\xi_2,\xi_3\}$ quantifying the extent of the polymer chain along its principal axes provide useful physical variables with which to color and interpret the low-dimensional diffusion map embeddings \cite{ferguson2010systematic, Rachael_2015, Ferguson_takens, Theodorou1985}. $\xi_1$ characterizes the stretching of the polymer along its longest axis, where large $\xi_1$ values correspond to linearly extended configurations. $\xi_2$ characterizes the width of the molecule, where large $\xi_2$ values indicate ``fat'' configurations possessing relatively large linear extensions perpendicular to the first principal axis. $\xi_3$ characterizes the thickness of the polymer orthogonal to the first and second principal axes; small $\xi_3$ values mean that the molecule is relatively planar, whereas large $\xi_3$ values mean that it is more globular.

\begin{figure*}[ht!]
\includegraphics[width=0.68\textwidth]{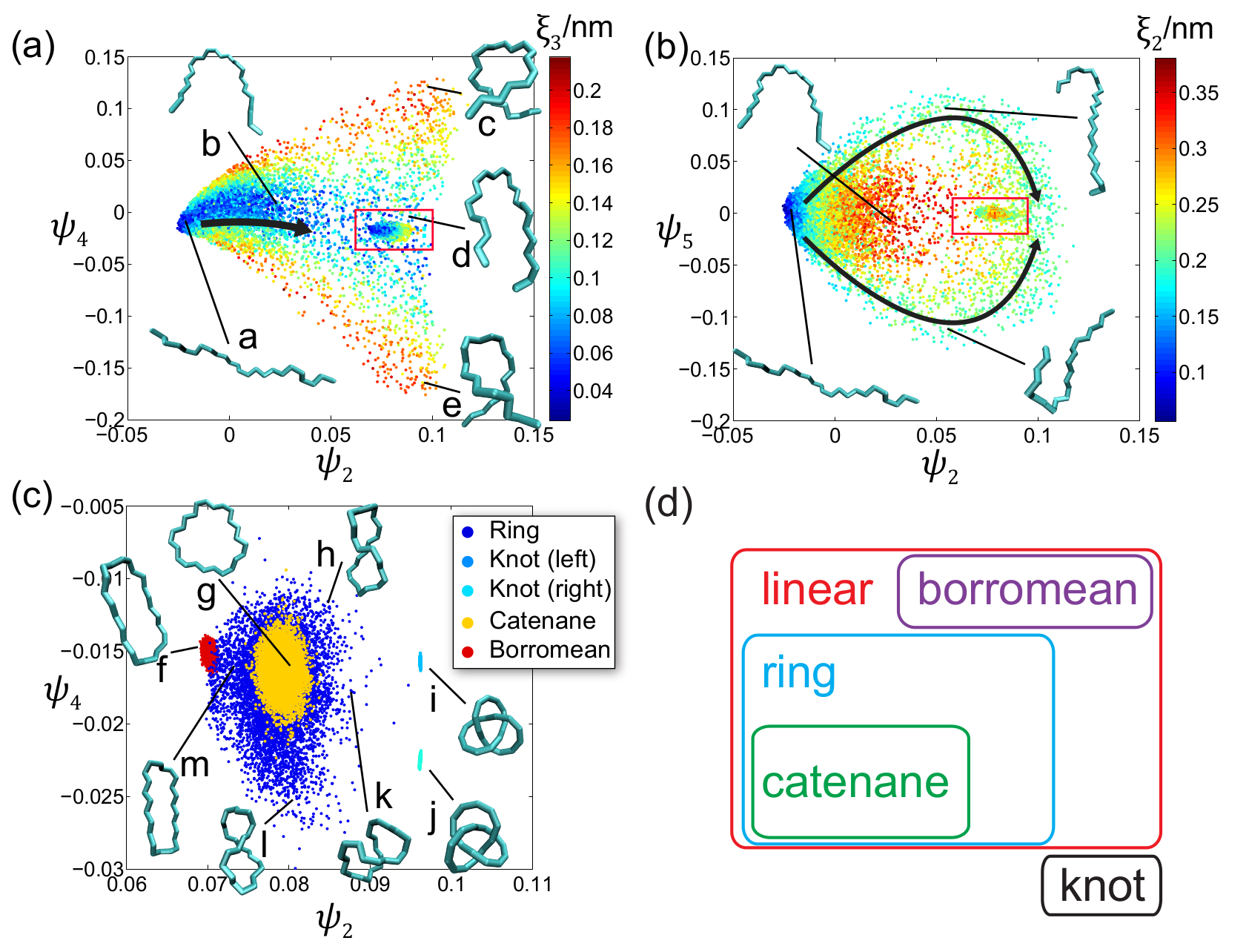}
\caption{\label{fig2} Projections of the 4D intrinsic manifold of the 24-mer polyethylene chain discovered by P-dMaps. Each point in the embedding corresponds to one of the chain configurations harvested from the molecular dynamics simulations conducted in each of the six topologies. Representative configurations are selected for visualization. (a) The embedding into $(\vec{\psi}_2, \vec{\psi}_4)$ reveals a triangular projection containing the large structural motions explored by the linear chain. Points are colored by the third principal moment of the gyration tensor $\xi_3$ quantifying the extent of the chain along its slimmest principal axis. The black arrows indicate the previously defined ``kink and slide'' collapse pathway for the linear chain \cite{ferguson2010systematic}. The region identified by a red box contains the subset of the manifold sampled by the rings and trefoil knots that are subject to far more restricted configurational motions relative to the linear chain by virtue of their molecular topologies. (b) The embedding into $(\vec{\psi}_2, \vec{\psi}_5)$ reveals the two ``kink and slide'' pathways that split depending on whether the kink in the asymmetric hairpin occur towards the head or the tail. Points are colored by the second principal moment of the gyration tensor $\xi_2$ quantifying the extent of the chain along its widest principal axis. (c) The $(\vec{\psi}_2, \vec{\psi}_4)$ embedding omitting the linear chain configurations reveals the distribution of the right and left-handed trefoil knots and ring topologies within the various supramolecular constructs. Points are colored by the supramolecular topology of the system. The rings populate a subset of the conformations adopted by the linear chain, whereas the knots populate a distinct region of the intrinsic manifold inaccessible to both the linear and ring topologies. (d) Schematic illustration of the relationship between the configurational ensembles sampled by the various chain topologies revealed by their distribution over the intrinsic manifold. The ring topologies populate a subset of the linear chain configurations, while the trefoil knots sample structurally distinct conformations. The configurations sampled by the homopolymer rings within catenane are nested within those of the isolated ring, while those within the Borromean ring stand distinct from both the isolated and catenane configurational ensembles.}
\end{figure*}

We present in \blauw{Fig.~\ref{fig2}a} the $(\vec{\psi}_2, \vec{\psi}_4)$ projection of the 4D intrinsic manifold of the 24-mer chains, where each point represents one of the 60,000 observed configurations. The manifold along this projection is approximately triangular in shape, and -- consistent with our previous work on linear polyethylene chains \cite{ferguson2010systematic, Rachael_2015, Ferguson_takens} -- this triangle encompasses the main structural motions of the linear 24-mer chain. The left vertex (configuration \textbf{a}) contains the approximately all-\textit{trans} extended chain configurations, whereas the two remaining vertices contain the left (\textbf{c}) and right (\textbf{e}) helices. The black arrow denotes the previously defined ``kink and slide'' pathway by which extended linear chains undergo hydrophobic collapse by the collective ejection of interstitial water molecules through an intermediate loose hairpin (\textbf{b}) to a tight symmetric hairpin with a dry core (\textbf{d}) \cite{ferguson2010systematic}. The bifurcation of this collapse pathway along an asymmetric hairpin possessing the kink near the head or tail of the linear molecule are more clearly illustrated in the $(\vec{\psi}_2, \vec{\psi}_5)$ projection in \blauw{Fig.~\ref{fig2}b}.

The ring and trefoil knot chain configurations are restricted to the region of the intrinsic manifold within the red rectangle in \blauw{Fig.~\ref{fig2}a,b}. The topological constraints introduced by the intramolecular covalent bond restricts these chains to sample a far smaller range of configurations than those accessible to the linear chain. In particular, the projection reveals them to be localized near the symmetric hairpin configurations of the linear chain with which they have the greatest structural similarity. Closer inspection reveals the configurations of the rings in isolation, catenane, and Borromean ring topologies (\blauw{Fig.~\ref{fig1}b,d,e}) exist within the region of the intrinsic manifold sampled by the linear chain, and therefore sample a subset of those configurations explored by the linear topology. Conversely, the projection of the right and left-handed trefoil knots (\blauw{Fig.~\ref{fig1}c}) lie outside of the region populated by the linear chain, indicating that the linear topology does not spontaneously adopt configurations proximate to the trefoil knot. 

To better resolve the distribution of the various ring/knot chain topologies over the manifold, we present in \blauw{Fig.~\ref{fig2}c} the $(\vec{\psi}_2, \vec{\psi}_4)$ projection of the manifold excluding the linear chains; projections into other combinations of the eigenvector pairs spanning the 4D intrinsic manifold are provided in \blauw{Figs.~S2-S5}. This projection illuminates the relationships between the configurational ensembles sampled by the chain as a function of supramolecular chemistry. The rings in catenane sample a subset of the configurations sampled by the ring in isolation, wherein interlocking with a second ring eliminates sampling of the elongated (\textbf{m}), twisted (\textbf{h},\textbf{l}), and boat (\textbf{k}) structures, restricting sampling to open and approximately circular ring configurations (\textbf{g}). Conversely, the individual rings constituting the Borromean ring exist at the very edge of the region populated by the isolated ring. Under the supramolecular topology of the Borromean ring wherein no pair of rings are interlocked but the triplet of rings is mechanically linked, the constituent rings are confined to a small region of the intrinsic manifold centered on elongated configurations (\textbf{f}). Finally, the left- (\textbf{i}) and right-handed (\textbf{j}) trefoil knots occupy distinct regions separated from the contiguous / nested regions populated by the isolated, catenane, and Borromean rings, but lie closest to the twisted (\textbf{h},\textbf{l}) and boat (\textbf{k}) ring structures. The trefoil can be formed from the twisted boat by cutting the ring, passing the central region of the chain between the split ends, and reconnecting the ends. Accordingly, these configurations are structurally proximate, but topologically distinct, and in reality one cannot interconvert into the other with our breaking intra-chain covalent bonds. A polymer chain with no excluded volume (i.e., a mathematically idealized 1D contour) in a ring topology could approach arbitrarily close to the trefoil knot by stacking the strands of the chain on top of each another but without one passing through the other as would be required to alter the topology from ring to knot. Our molecular model of the chain does possess excluded volume that accounts for the gap between the ring and knot configurational ensembles on the manifold. 

In sum, our application of composite P-dMaps revealed the relationships between the configurational ensembles of the chain in various supramolecular chemistries by comparing the populated regions of the intrinsic manifold. A schematic illustration of these relationships is presented in \blauw{Fig.~\ref{fig2}d}.

\begin{figure*}[ht!]
\includegraphics[width=0.99\textwidth]{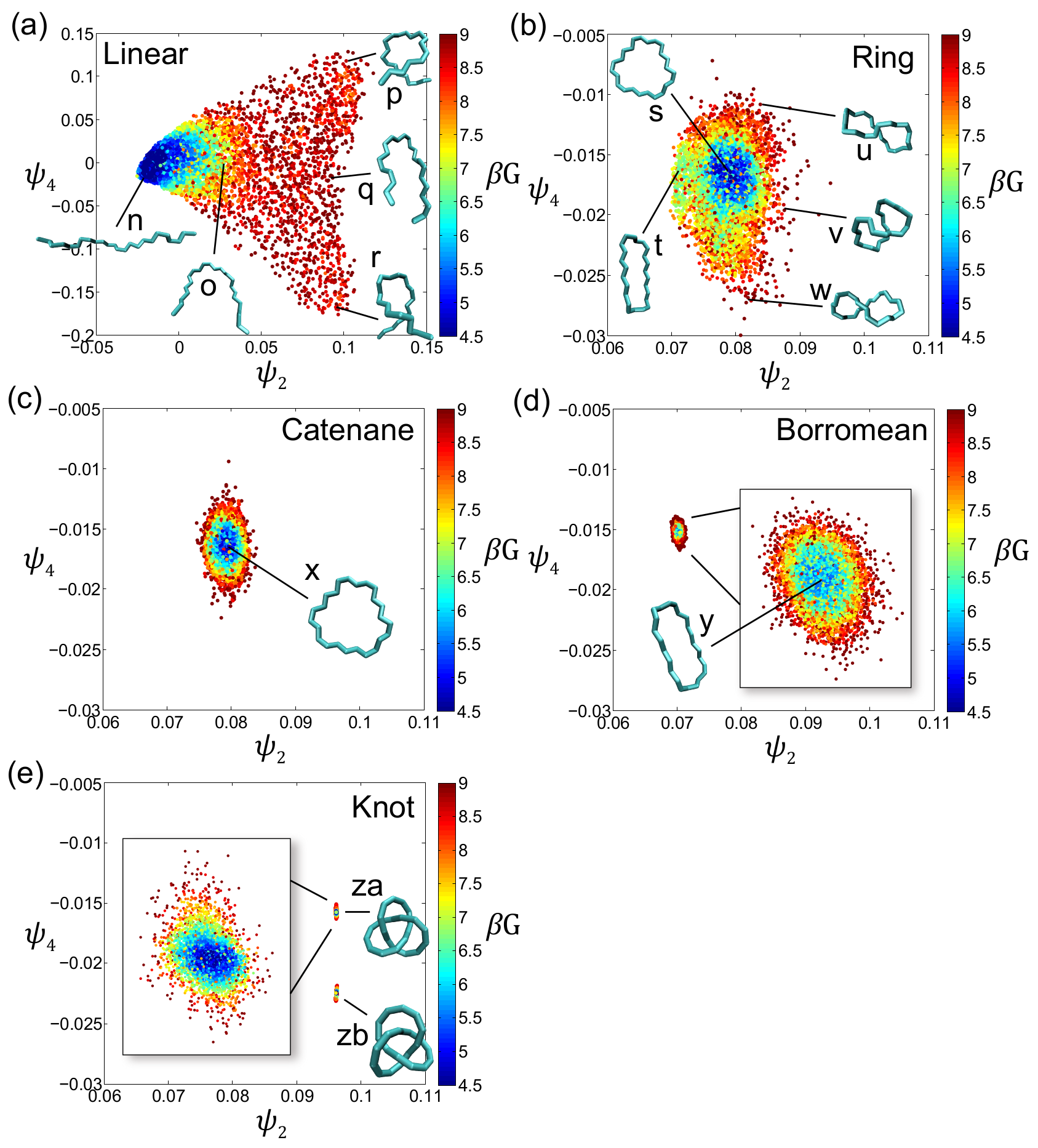}
\caption{\label{fig3} Free energy surfaces for 24-mer polyethylene chain in each of the six supramolecular topologies projected into $(\vec{\psi}_2, \vec{\psi}_4)$ under different geometric constraints. Free energy surfaces for (a) the linear chain, (b) the isolated ring, (c) the ring in catenane, (d) the ring in Borromean, (e) the trefoil knot.}
\end{figure*}

\textbf{Free energy surfaces}. We present in \blauw{Fig.~\ref{fig3}} the free energy surface over the intrinsic manifold discovered by P-dMaps for the polymer chain in each of the six supramolecular topologies studied in this work. For representational convenience, we present the 2D projections of these 4D landscapes into $(\vec{\psi}_2, \vec{\psi}_4)$. Projections into other eigenvector pairs are provided in \blauw{Figs.~S6-S10}.

The FES of the isolated linear chain in \blauw{Fig.~\ref{fig3}a} is consistent with our previous work, exhibiting a global minimum at the all-\textit{trans} configuration (\textbf{n}) that is in equilibrium with slightly less stable loose hairpins (\textbf{o}) and tight hairpins (\textbf{q}) and hydrophobically-stabilized left- and right-handed helices (\textbf{p},\textbf{r}) that lie $\sim$4 $k_BT$ higher in free energy. 

Considering now the ring polymers within different supramolecular environments, the FES of the isolated ring \blauw{Fig.~\ref{fig3}b} exhibits a global free energy minimum at the center of the manifold corresponding to the approximately circular open ring (\textbf{s}). To the left of the manifold lie elongated elliptical rings (\textbf{t}) that lie only $\sim$2 $k_BT$ higher in free energy indicating that they are also highly favorable stable structures. To the top and bottom of the manifold lie the right- and left-handed twisted (i.e., figure-8) structures (\textbf{u},\textbf{w}), and boat configuration (\textbf{v}) that lie $\sim$4 $k_BT$ higher in free energy. Accordingly, these structures -- while still thermally accessible -- are less favorable than the circular and elliptical ring structures. 

The rings within the catenane topology in \blauw{Fig.~\ref{fig3}c} possess a FES with the same global minimum as the isolated ring centered on the approximately circular configuration (\textbf{x}), but the diversity of accessible configurations is much smaller. The FES is essentially parabolic indicating a Gaussian probability distribution centered on the global minimum with sampling around this state driven by thermal fluctuations. Absent are the elongated, twisted, and boat configurations that are disfavored by the interlocked topology of catenane that prevents the ring from accessing these structures. 

The FES of the homopolymer rings within the Borromean ring in \blauw{Fig.~\ref{fig3}d} lies to the left edge of the isolated ring landscape, centered on the elongated ring structure (\textbf{y}). Again, the probability distribution is approximately Gaussian. It is challenging to see in the $(\vec{\psi}_2, \vec{\psi}_4)$ projection, but the FES of the isolated and Borromean rings do not intersect, populating distinct regions of the intrinsic manifold that is more clearly resolved within projections of the FES into the other eigenvector pairs presented in \blauw{Fig.~S5}. The effect of the Borromean ring supramolecular topology is to force the constituent rings to adopt configurations that are not sampled at thermal equilibrium by the homopolymer ring in isolation. It is difficult to distinguish the differences between the elongated configurations sampled by the isolated (\textbf{t}) and Borromean rings (\textbf{y}) by visual inspection alone, but careful analysis reveals the latter to be narrower and more elongated with more strained intramolecular bond angles. These differences are revealed by the P-dMap embedding in the gap between the FES in the intrinsic manifold, and we explore the differences further through an energetic analysis reported in the \blauw{next section}.

The trefoil knot FES is located to the right of the intrinsic manifold, separated from one another by their left- (\textbf{za}) and right-handed (\textbf{zb}) chirality and from the linear and ring configurational ensembles. The very restrictive structural constraints imposed by the trefoil knot topology on these short 24-mer chains makes these structures very rigid, and the FES is tightly localized on the global minimum with Gaussian configurational fluctuations driven by thermal noise.

\textbf{Potential energy breakdown}. To gain further insight into the chain behavior under the various supramolecular topologies, we report  the various contributions to the potential energy of the 24-mer chain in each supramolecular topology in \blauw{Table~\ref{tb1}}. The intramolecular contributions comprise angle bending, dihedral rotation, and intramolecular Lennard-Jones interactions between the united atom beads constituting the chain. There are no bond stretching contributions since the bond lengths are rigid in the TraPPE potential \cite{Martin1998}. Intermolecular contributions comprise exclusively the Lennard-Jones interactions. There are no Coulombic contributions as all united atoms are uncharged.

\begin{table}[ht!]
  \caption{Potential energy breakdown for the 24-mer polyethylene chain within the various supramolecular topologies. Quantities for the trefoil knot are averaged over both the left- and right-handed topologies. Mean values are computed by averaging over the 100 ns molecular simulation trajectory, and standard errors computed by block averaging over five 20 ns time blocks. Values are reported in kJ/mol.}
\label{tb1}
  \begin{tabular}{| c | c | c | c | c | c | c | c | c |}
    \hline
    Topology & Angle & Dihedral & Intra LJ  & Inter LJ & Total LJ & Total \\
    \hline
    Linear & 27.75 & 52.08 & -12.71 & -166.69 & -179.40 & -99.57 \\
    & $\pm$ 8.43 & $\pm$ 11.37 & $\pm$ 4.59 & $\pm$ 16.74 & $\pm$ 15.04 & $\pm$ 20.65 \\
    \hline
    Isolated & 30.26 & 64.67 & -17.95 & -140.71 & -158.67 & -63.74\\
    ring & $\pm$ 8.75 & $\pm$ 11.85 & $\pm$ 3.41 & $\pm$ 12.35 & $\pm$ 11.57 &  $\pm$ 18.73\\    
    \hline
    Trefoil & 1552.00	& 360.00 & 25829.00 & -130.00 & 25699.00 & 27611.00 \\
    knot & $\pm$ 5.01 & $\pm$ 0.97 & $\pm$ 106.54 & $\pm$ 101.24 & $\pm$ 34.34 & $\pm$ 34.72 \\
    \hline
    Catenane & 29.53 & 58.86 & -15.09 & -160.75 & -175.85 & -87.46\\
    & $\pm$ 8.48 & $\pm$ 10.54 & $\pm$ 1.06 & $\pm$ 10.37 & $\pm$ 10.38 & $\pm$ 17.05\\
    \hline
    Borromean & 327.43 & 126.74 & -13.41 & 653.20 & 639.79 & 1094.00 \\
    ring & $\pm$ 28.78 & $\pm$ 8.81 & $\pm$ 0.21 & $\pm$ 37.19 & $\pm$ 37.10 & $\pm$ 47.77\\
    \hline
  \end{tabular}
\end{table}

The angle and dihedral contributions are $\sim$30 kJ/mol and 50-60 kJ/mol, respectively, for the linear, isolated ring, and catenane topologies, indicating the relatively unstrained conformations adopted by the chain in these systems. This stands in contrast to the Borromean ring and trefoil knot, which, respectively, possess dihedral contributions that are 2-5 times larger and angle contributions that are 1-2 orders of magnitude larger. This reflects the strong perturbations of the chain structure induced by the strained topology of the trefoil knot, and the strong confining effects of the two other chains in the Borromean ring.

The intramolecular Lennard-Jones contributions modeling the dispersion interactions between monomers within the chain are $\sim$(-15) kJ/mol for the linear, isolated ring, catenane, and Borromean ring topologies, indicating that intra-chain dispersion interactions are relatively unperturbed in all four of these topologies. The trefoil knot stands out with an intramolecular Lennard-Jones contribution of almost 26,000 kJ/mol. The source of this large positive term is the exceedingly strained configuration imposed by the trefoil topology on such a short chain, such that the crossed strands of the chain bring the constituent atoms into such close proximity that they interact via hard core repulsions. Given that the approximate formation energy of a \ce{C-C} bond is 346 kJ/mol \cite{haynes2016crc, Benson1965}, one should expect the trefoil knot for the 24-mer chain to be energetically unstable and that this object could not be chemically synthesized. We are at liberty, however, to simulate this hypothetical object using a forcefield that does not allow for covalent bond scission, and it stands as a useful comparison to the 50-mer trefoil knot that by virtue of its much longer chain length is far more relaxed within the trefoil and possesses a far more reasonable intramolecular Lennard-Jones energy (cf.\ \blauw{Table~\ref{tb2}}).

The intermolecular Lennard-Jones interactions of each ring homopolymer with the surrounding solvent and any other chains within the supramolecular construct are all comparable for the linear, isolated ring, trefoil knot, and catenane topologies at (-130)-(-170) kJ/mol. For a chain in the Borromean ring, however, this contribution is large and positive at $\sim$650 kJ/mol. The situation here is analogous to the intramolecular Lennard-Jones contributions for the trefoil knot, but where in this case the hard core interactions are induced between atoms in different chains due to the stringent topological constraints of the Borromean ring for a 24-mer polymer chain. Accordingly, we expect that this chemistry should also be chemically unstable, but again it provides a useful comparison for the 50-mer Borromean ring where the interaction energy of this more relaxed topology is more reasonable.

\subsection*{3.2 Structure, energy, and free energy surfaces for 50-mer chains}



\textbf{Configurations}. Similar to the 24-mer chain, we analyzed the 60,000 chain configurations of the 50-mer from 100 ns molecular dynamics simulations in each of the six topologies illustrated in \blauw{Fig.~\ref{fig1}}. A gap in the eigenvalue spectrum after eigenvalue $\lambda_4$ informed the construction of 3D nonlinear embedding in the top three non-trivial eigenvectors $( \vec{\psi}_2, \vec{\psi}_3, \vec{\psi}_4)$ (\blauw{Eqn.~\ref{eqn:dMapEmbed}}, \blauw{Fig.~S1b}). Similar to our previous work, we find that two of these non-trivial eigenvectors, $\vec{\psi}_2$ and $\vec{\psi}_3$, are functionally dependent \cite{ferguson2010systematic}. This is clearly evinced in a scatter plot of $\vec{\psi}_2$ against $\vec{\psi}_3$ that reveals a narrow belt-like structure (\blauw{Fig.~S11}). Since the width of the belt is small relative to its arclength, $\vec{\psi}_3$ is effectively slaved to $\vec{\psi}_2$ and provides no substantial new information in the low-dimensional projection. Physically, this corresponds to two diffusion map eigenvectors that characterize the same collective dynamical mode \cite{ferguson2010systematic}. Accordingly, we can eliminate $\vec{\psi}_3$ from our diffusion map embedding without any significant loss of information, and we are at liberty to construct much more easily representable and interpretable 2D diffusion map embeddings into $(\vec{\psi}_2, \vec{\psi}_4)$. Again, we employ the principal moments of the gyration tensor $(\xi_1,\xi_2,\xi_3)$ to color and interpret the embeddings. 

\begin{figure*}[ht!]
\includegraphics[width=0.84\textwidth]{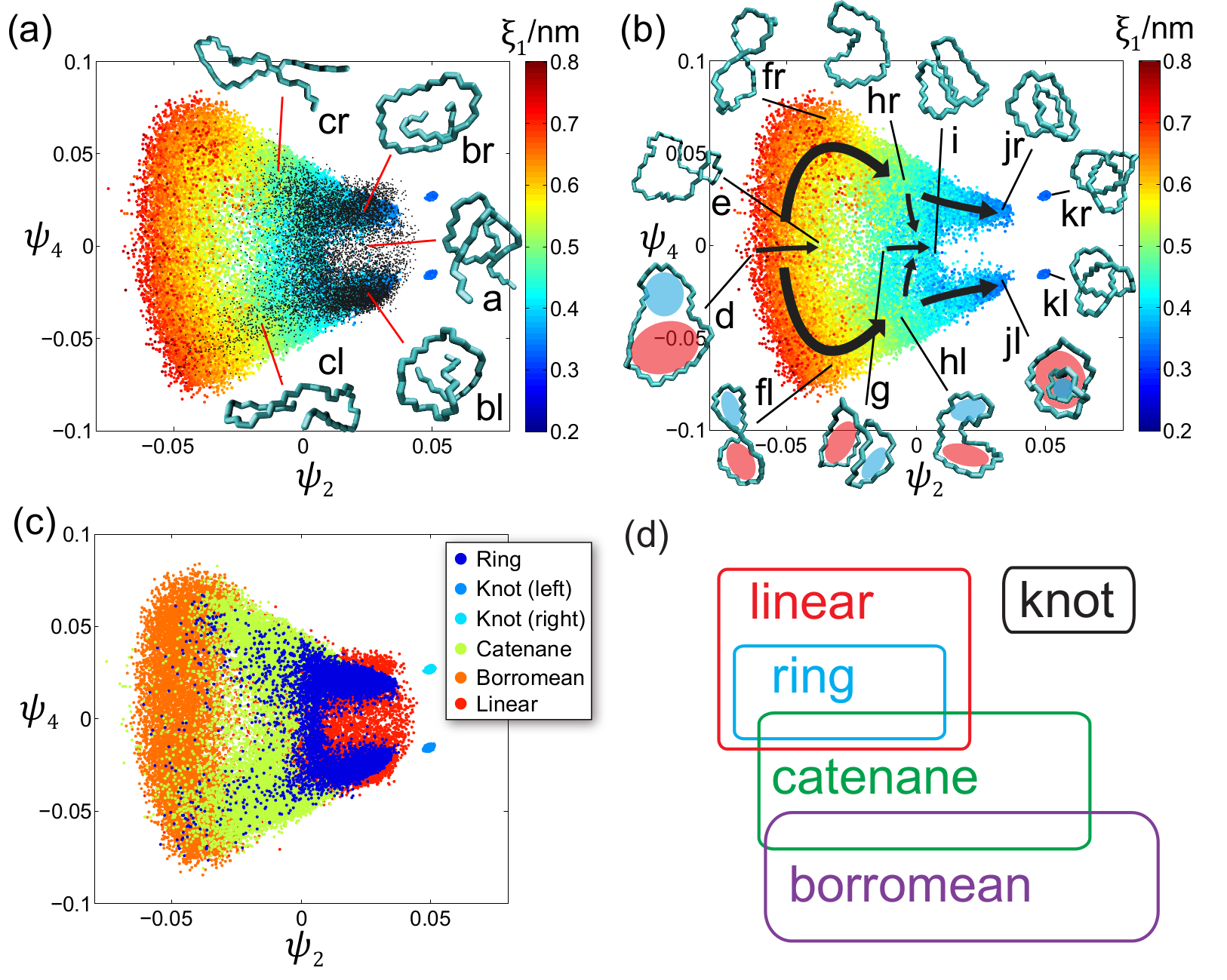}
\caption{\label{fig4} Projection of the effectively 2D intrinsic manifold of the 50-mer polyethylene chain discovered by P-dMaps into ($\vec{\psi}_2$, $\vec{\psi}_4$). Each point in the embedding corresponds to one of the chain configurations harvested from the molecular dynamics simulations conducted in each of the six topologies. Representative configurations are selected for visualization. (a) Points corresponding to ring molecules are colored by the first principal moment of the gyration tensor $\xi_1$. Points corresponding to linear chains are colored in black to differentiate these from the ring configurations, and are not represented in the heat map defined by the color bar. (b) Reproduction of panel a, but omitting the linear chain configurations to better illustrate the embedding of the ring topologies. Points are colored by $\xi_1$. Black arrows indicate the folding pathways of the ring from the open state to the collapsed state. The red and blue ellipsoids show the planes containing the united atoms comprising the two ends of the ring, and help illustrate molecular folding. (c) Reproduction of panel a, but with points colored by the supramolecular topology. (d) Schematic illustration of the relationship between the configurational ensembles sampled by the various chain topologies revealed by their distribution over the intrinsic manifold. The isolated ring populate a subset of the conformations adopted by the linear chain. Rings in catenane system partially overlap with the isolated ring and the linear chain. Rings in Borromean system partially overlap with catenane. The trefoil knots populate a distinct region of the intrinsic manifold inaccessible to the other systems.}
\end{figure*}

The projection of the 60,000 points into $(\vec{\psi}_2, \vec{\psi}_4)$ are presented in \blauw{Fig.~\ref{fig4}a}. The embedding illustrates that $\vec{\psi}_2$ and $\xi_1$ are inversely correlated, such that elongated molecular configurations reside at small $\vec{\psi}_2$, and vice-versa. In contrast to the 24-mer chain where the ring topologies populated a subset of the intrinsic manifold sampled by the linear chain, the 50-mer linear chain is restricted to a region on the right edge of the manifold corresponding to relatively collapsed chain configurations. Representative snapshots of the linear chain illustrating the right- and left-handed helices (configurations \textbf{br}, \textbf{bl}), and right- and left-handed twisted hairpins (\textbf{cr}, \textbf{cl}). The regions of the linear chain embedding from which these configurations are drawn overlap with the region populated by the ring topologies, indicating that these representative chain configurations have approximate analogs in the ring systems. Conversely, the coiled linear chain configuration (\textbf{a}) comes from a region not populated by the ring topologies, and is therefore inaccessible to the rings.

A reproduction of the projection omitting the linear chain configurations is shown in \blauw{Fig.~\ref{fig4}b}, and each point colored with $\xi_1$. Analogous plots but with points colored by $\xi_2$ and $\xi_3$ are provided in \blauw{Fig.~S12}. Neglecting for a moment the fact that the points were drawn from five different ring supramolecular topologies, we first consider the diversity of and relation between the various sampled ring configurations. Starting from the left side of the manifold corresponding to open ring configurations (\textbf{d}), we see that the molecule can fold into a boat configuration (\textbf{e}) by following the linear path along $\vec{\psi}_4$ = 0. Further collapse proceeds by continuing along this path to (\textbf{g}) and then (\textbf{i}) corresponding to tightening of the boat topology. We indicate this path by narrow black arrows. We omit the arrow between (\textbf{e}) and (\textbf{g}) since, as we discuss below, there is a gap in the manifold here, and this transition is not actually sampled.

Alternatively, the open ring configuration at (\textbf{d}) can progress along either of the two thick curved arrows corresponding to the formation of twisted figure-8 configurations with either right-handed (\textbf{fr}) or left-handed (\textbf{fl}) helicity. The two sides of the ring approximately reside in the planes illustrated by the red and blue ellipses projected onto the left-handed representative structures. These are initially approximately co-planar in (\textbf{d}), but then fold over one another in the figure-8 shapes (\textbf{fr}, \textbf{fl}), and ultimately stack into an open and twisted boat-like configuration (\textbf{hr}, \textbf{hl}). These twisted boats can then relax their twist to fold into the planar boats in (\textbf{g}, \textbf{i}), or further twist so that the planes stack in the opposite sense to form the collapsed twisted boats in (\textbf{jr}, \textbf{jl}).

The right-handed (\textbf{kr}) and left-handed (\textbf{kl}) trefoil knot configurations populate the right edge of the manifold. Again these configurations are projected into islands separated from the mainland of the intrinsic manifold populated by the linear chain and rings. As discussed above for the 24-mer chain, this separation arises since transformation into the knot configurations from the ring would require breaking of covalent bonds, and from the linear chain a highly unfavorable and statically constrained threading of the chain.

Returning to the relationships between the various supramolecular topologies over the manifold, we present a projection of the embedding in \blauw{Fig.~\ref{fig4}c} in which we have colored the points by their supramolecular topology. This plot reveals an interesting series of overlapping relationships between the configurational ensembles sampled in each of the supramolecular topologies that we illustrate schematically in \blauw{Fig.~\ref{fig4}d}. Ring polymers in the Borromean ring occupy the most expanded configurations on the left of the manifold. Catenane rings occupy the middle of the manifold, overlapping with the Borromean rings to the left, and isolated rings and linear chains to the right. The isolated ring occupies a subset of the configurations populated by the linear chain. The right and left trefoil knots are located away from region of the manifold populated by the other supramolecular systems. By analyzing the simulation trajectories of the six supramolecular topologies, P-dMaps has uncovered the configurational ensembles populated by each system and their relative overlaps and relationships.

\begin{figure*}[ht!]
\includegraphics[width=0.95\textwidth]{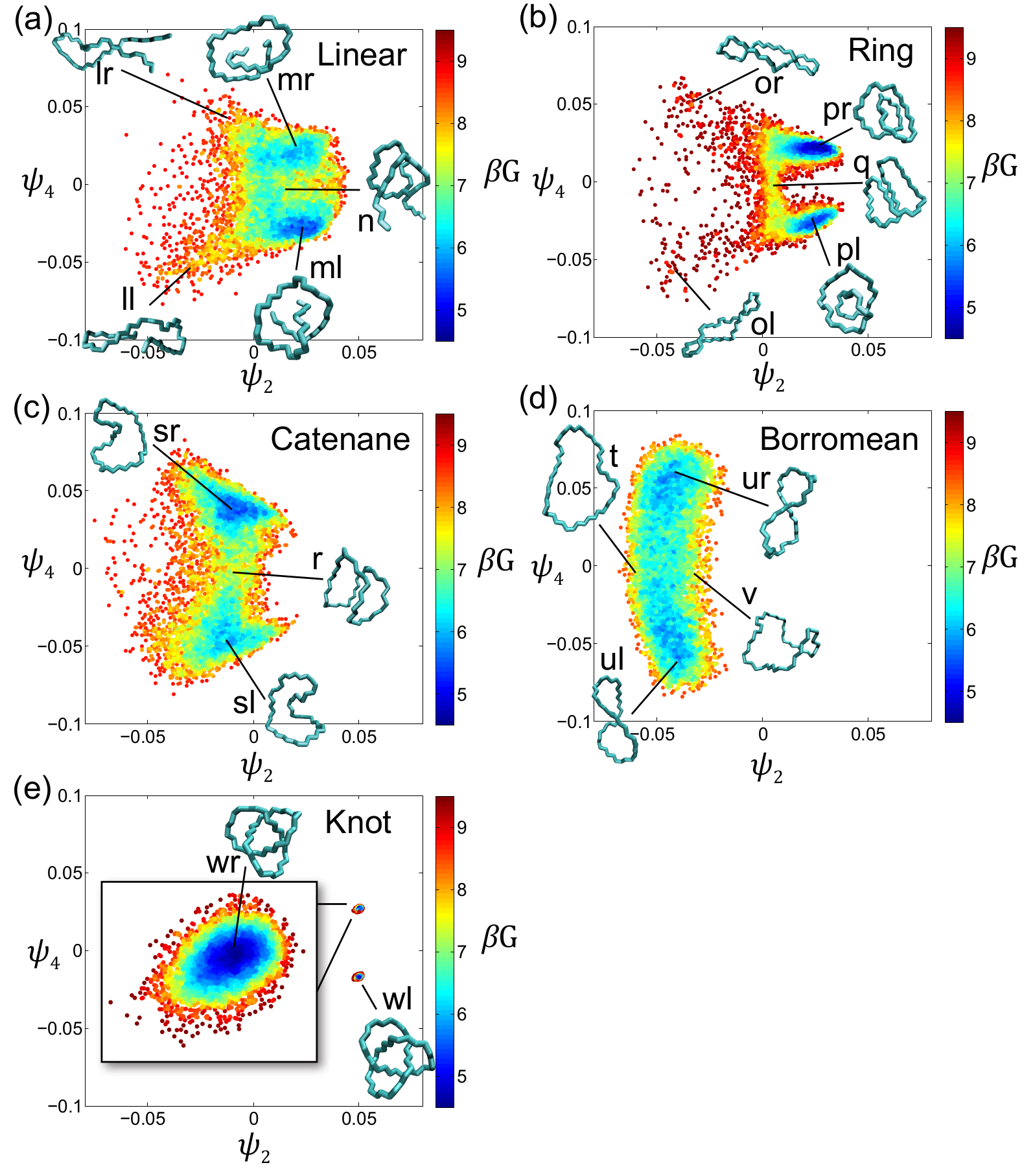}
\caption{\label{fig5} Free energy surfaces for 50-mer polyethylene chain in each of the six supramolecular topologies projected into $(\vec{\psi}_2, \vec{\psi}_4)$ under different geometric constraints. Free energy surface for (a) the linear chain (b) the isolated ring (c) the ring in catenane  (d) the ring in Borromean  (e) the trefoil knot.}
\end{figure*}

\textbf{Free energy surfaces}.We present in \blauw{Fig.~\ref{fig5}} the free energy surface over the intrinsic manifold spanned by $(\vec{\psi}_2, \vec{\psi}_4)$ discovered by P-dMaps for the 50-mer polymer chain in each of the six supramolecular topologies.

The FES for the linear 50-mer linear chain is shown in \blauw{Fig.~\ref{fig5}a}. There are two minima containing the right-handed (\textbf{mr}) and left-handed (\textbf{ml}) helices, which represent the most stable states of the chain. A low-free energy pathway connects the two minima containing the coiled structure (\textbf{n}). The small free energy barrier of $\sim$1.5 $k_B T$ indicates that the left- and right-handed configurations can easily interconvert. The left side of the manifold contains the more elongated right-handed (\textbf{lr}) and left-handed (\textbf{ll}) twisted hairpin structures. The free energy of these configurations is $\sim$4 $k_B T$ higher than the global minima, indicating the relative unfavorability of these structures compared to the hydrophobically collapsed helices \cite{ferguson2010systematic,chakrabarty2009self,underwood2010long}.

The FES for the isolated ring simulation is shown in \blauw{Fig.~\ref{fig5}b}. Again we observe two minima corresponding to the right-handed (\textbf{pr}) and left-handed (\textbf{pl}) collapsed twisted boats. The minima are linked by a low-free energy pathway containing the collapsed boat configuration (\textbf{q}) and possessing a free energy barrier of $\sim$3.0 $k_B T$. The topological constraints of the ring make this interconversion pathway both narrower and higher than the analogous pathway for the linear chains. The tails to the left of the manifold contain the left-handed (\textbf{ol}) and left-handed (\textbf{or}) twisted rings that lie $\sim$4 $k_B T$ higher than the global minima. As stated above, the ring polymers adopt a subset of the configurational ensemble populated by the linear chains. The free energy surfaces of these two systems are remarkably similar, with the only significant difference being the presence of a gap between the two lobes containing the two minima in the ring topology due to the topological constraints introduced by joining the free ends of the chain. It is surprising that the linear and ring topologies should share so much similarity in their configurational exploration and relative stability of the various chain configurations.

The FES for the ring molecule in catenane is shown in \blauw{Fig.~\ref{fig5}c}. The populated region of the manifold is shifted left relative to the linear chain and isolated rings, meaning that ring configurations tend to be more open and extended. This is a consequence of the interlocking of the two rings that introduces steric constrains that preclude full hydrophobic collapse. The projection of the catenane configurational ensemble partially overlaps with that of the linear chain and isolated ring, indicating that the most compact configurations of the former are sampled by the most expanded configurations of the latter two. Again we observe two minima corresponding to the right-handed (\textbf{sr}) and left-handed (\textbf{sl}) semi-folded structures. These configurations cannot fully collapse due to the presence of an interlocking partner ring. These two structures can interconvert via a boat configuration (\textbf{r}) located at the top of a $\sim$2 $k_B T$ free energy barrier.

The FES for ring polymers in the Borromean ring are shown in \blauw{Fig.~\ref{fig5}d}. These points are located at the extreme left of the composite manifold meaning that these rings are the most expanded due to the mutual threading of the rings within this supramolecular assembly. There is some overlap of the Borromean ring and catenane configurational ensembles. Interestingly, the two minima in the Borromean ring correspond not to collapsed configurations, but twisted figure-8 structures with left-handed (\textbf{ul}) and right-handed (\textbf{ur}) chiralities. They reside at the bottom of weak minima within a superbasin in the FES, and are connected by a untwisting/twisting pathway with a barrier of $\sim$1 $k_B T$. The left of the populated region contains very open and expanded rings (\textbf{t}) and the right asymmetrically twisted rings (\textbf{v}). As mentioned above, further collapse of the (\textbf{v}) configuration by moving right along the $\vec{\psi}_4$ = 0 contour is prohibited by the geometric constraints of the Borromean ring.

Finally, the FES of the trefoil knots are shown in \blauw{Fig.~\ref{fig5}e}. The high rigidity of these structures mean that the configurational ensemble is essentially Gaussian distributed around the most stable configuration and the resulting FES is parabolic.

\textbf{Potential energy breakdown}. We report in \blauw{Table.~\ref{tb2}} the breakdown of the contributions to the potential energy of the 50-mer chain as a function of supramolecular topology. The angle and dihedral contributions are $\sim$60 kJ/mol and $\sim$130 kJ/mol, respectively, for all systems except the trefoil knot, which possesses slightly larger values of $\sim$ 87 kJ/mol and $\sim$ 200 kJ/mol. Although these energetic contributions for the knot are substantially larger than for the other systems, the magnitude of the difference is far smaller than for the 24-mer chain, reflecting that fact that the longer chain is less strained and more stable in the knot configuration (cf.\ \blauw{Table.~\ref{tb1}}).

\begin{table}[ht!]
  \caption{Potential energy breakdown for the 50-mer polyethylene chain within the various supramolecular topologies. Quantities for the trefoil knot are averaged over both the left- and right-handed topologies. Mean values are computed by averaging over the 100 ns molecular simulation trajectory, and standard errors computed by block averaging over five 20 ns time blocks. Values are reported in kJ/mol.}
\label{tb2}
  \begin{tabular}{| c | c | c | c | c | c | c | c | c |}
    \hline
    Topology & Angle & Dihedral & Intra LJ  & Inter LJ & Total LJ & Total \\
    \hline
    Linear & 60.75 & 126.24 & -79.06 & -231.49 & -310.55 & -123.56 \\
    & $\pm$ 12.44 & $\pm$ 15.96 & $\pm$ 10.16 & $\pm$ 20.34 & $\pm$ 17.75 & $\pm$ 26.92 \\
    \hline
    Isolated & 63.31 & 137.94 & -81.45 & -224.78 & -306.24 & -104.99\\
    ring & $\pm$ 12.68 & $\pm$ 16.25 & $\pm$ 8.82 & $\pm$ 16.56 & $\pm$ 15.42 &  $\pm$ 25.74\\    
    \hline
    Trefoil & 87.12	& 206.41 & 37.27 & -211.07 & -173.80 & 119.73 \\
    knot & $\pm$ 16.14 & $\pm$ 22.70 & $\pm$ 16.82 & $\pm$ 12.21 & $\pm$ 20.57 & $\pm$ 34.63 \\
    \hline
    Catenane & 63.06 & 130.96 & -48.41 & -309.91 & -358.33 & -164.31\\
    & $\pm$ 12.71 & $\pm$ 16.28 & $\pm$ 8.02 & $\pm$ 18.78 & $\pm$ 16.25 & $\pm$ 26.28\\
    \hline
    Borromean & 62.76 & 127.81 & -33.30 & -343.68 & -376.99 & -186.42 \\
    ring & $\pm$ 12.45 & $\pm$ 15.94 & $\pm$ 4.29 & $\pm$ 16.60 & $\pm$ 16.16 & $\pm$ 25.89\\
    \hline
  \end{tabular}
\end{table}

The intramolecular Lennard-Jones contributions for all systems except the knot are favorable and lie in the range (-30)-(-85) kJ/mol. The constrained topology of the trefoil knot is $\sim$37 kJ/mol, which -- although unfavorable -- is three orders of magnitude smaller than that for the 24-mer knot. The intermolecular Lennard-Jones interactions are all large and negative in the range (-200)-(-350) kJ/mol. This result for the Borromean ring stands in contrast to that for the 24-mer chain, which was large and positive at $\sim$640 kJ/mol. The 50-mer chain in the Borromean ring is far more relaxed and stable by virtue of the larger chain length.

\subsection*{3.3 Molecular and supramolecular size and center of mass separation}

\begin{figure*}[ht!]
\includegraphics[width=0.7\textwidth]{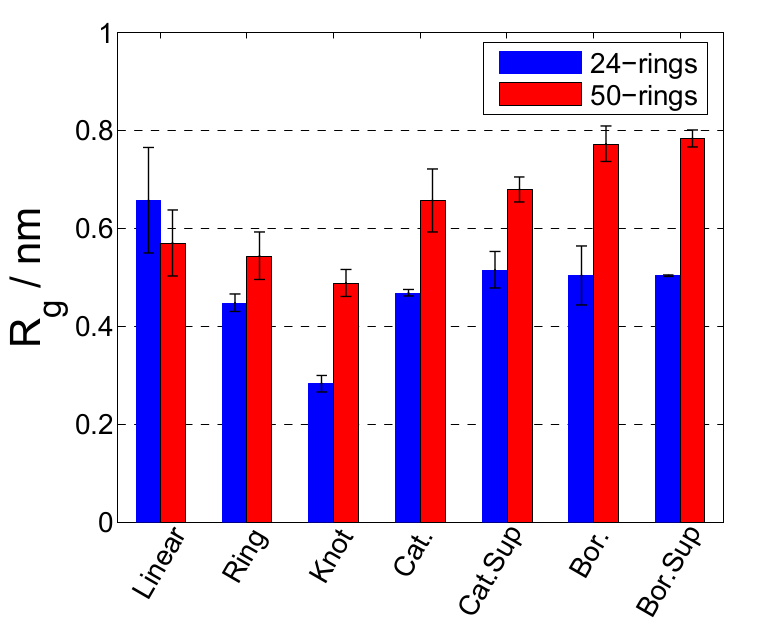}
\caption{\label{fig6} Radius of gyration of a single homopolymer chain in the linear, ring, trefoil knot, catenane, and Borromean ring topologies, and of the 2-ring catenane supramolecular complex and 3-ring Borromean ring complex. Mean values are computed over the 100 ns molecular dynamics trajectories (for the trefoil knot, the left- and right-handed systems are averaged), and error bars computed by block averaging over five 20 ns blocks.}
\end{figure*}

\textbf{Molecular and supramolecular radius of gyration}. To quantify the size of the 24-mer and 50-mer polymer chains as a function of supramolecular topology, we computed the single molecule radius of gyration (\blauw{$R_g$}) of the chain in each environment. We also computed the supramolecular radius of gyration of the 2-ring catenane complex and 3-ring Borromean ring complex at each degree of polymerization. Results are presented in \blauw{Fig.~\ref{fig6}}.

In all cases the 24-mer chain has a smaller $R_g$ than the 50-mer chain as might be expected from its smaller degree of polymerization. This trend is inverted, however, in the case of the isolated linear chain where $R_g$ = 0.65 nm for the 24-mer chain compared to $R_g$ = 0.56 nm for the 50-mer, although these values are indistinguishable within error bars. Work by Livadaru and Kovalenko who showed the radius of gyration of polyethylene in water increases very slowly with degree of polymerization due to the hydrophobic collapse and packing of the chain in poor solvent \cite{Lucian2004, Lucian2005}. The indistinguishability of $R_g$ within error bars for the 24-mer and 50-mer linear chains is consistent with this slow increase reported in that work. As chain length increases further, we would expect the $R_g$ dependence to enter the ideal scaling regime where $R_g \sim N^{1/3}$.


For both the 24-mer and 50-mer chains the ordering of the single chain $R_g$ in the various supramolecular topologies follows the same trend, with trefoil knot $<$ isolated ring $<$ catenane $<$ Borromean ring. That the knot is the smallest is unsurprising due to its extremely compact topology, and the increasing size of the ring in the remaining topologies is a consequence of the introduction of supramolecular constraints that favor more elongated and narrow ring configurations. This rank ordering is more pronounced for the longer 50-mer chain compared to the shorted 24-mer. The size of the 2-ring catenane supramolecular assembly is slightly larger than that of a single ring within that assembly by $\Delta R_g$ = 0.05 nm for the 24-mer chain, and $\Delta R_g$ = 0.02 nm for the 50-mer. This small but measurable increase is due to the fact that the interlocked catenane rings do not share a center of mass, so that the distribution of mass around the center of the 2-ring complex is more dispersed than that for a single ring. Conversely, all three constituent homopolymer rings in the concentric topology of the Borromean ring do share a center of mass, and there is no measurable difference in $R_g$ between the 3-ring Borromean complex and a single constituent homopolymer ring.

\begin{figure*}[ht!]
\includegraphics[width=0.99\textwidth]{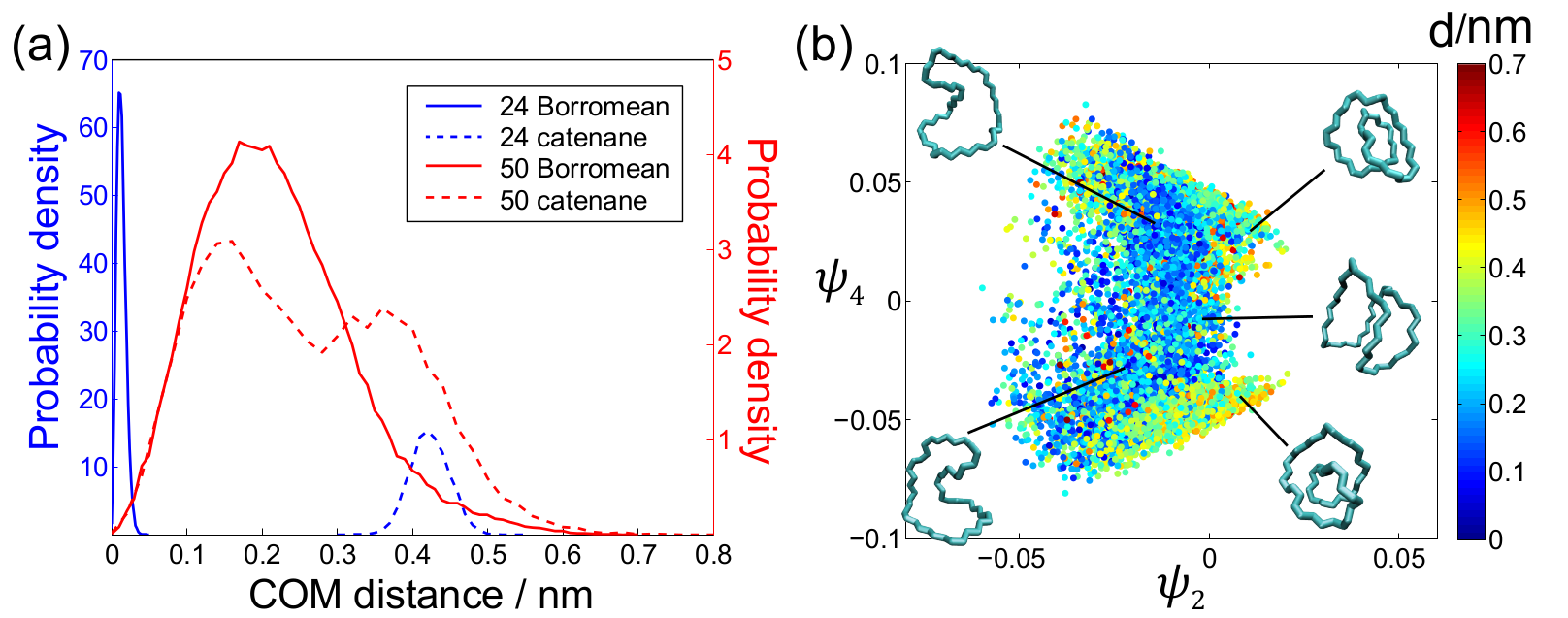}
\caption{\label{fig7} Pairwise center of mass separation of the rings constituting the catenane and Borromean ring supramolecular assemblies. (a) Probability density distribution of the center of mass distance between pairs of rings in the catenane and Borromean ring supramolecular topologies for the 24-mer and 50-mer chains. (b) Projection of the 50-mer catenane single ring configurations onto the intrinsic manifold projection into $(\vec{\psi}_2,\vec{\psi}_4)$ and colored by the center of mass distance between the two rings comprising the assembly.}
\end{figure*}

\textbf{Supramolecular center of mass separation}. To quantify the center of mass (\blauw{COM}) separation between the rings constituting the catenane and Borromean ring supramolecular complexes, we computed from our molecular simulations the probability distribution of the pairwise COM separations for the 24-mer and 50-mer chains that we present in \blauw{Fig.~\ref{fig7}a}. For the 24-mer chain, the Borromean ring pairwise COM is a tight distribution centered close to zero that is well fit by a Gaussian centered at a COM separation of 0.01 nm. Conversely, the 24-mer chain catenane topology probability distribution is removed to much larger COM separations and is well fit by a Gaussian centered on 0.43 nm. For the 50-mer chains, the probability distribution for COM separations in the Borromean ring is a much broader non-Gaussian distribution with a mode of 0.2 nm. The higher degree of polymerization provides these rings with greater intramolecular flexibility relative to the 24-mer chains, as illustrated in the larger range of configurations sampled by the longer chains (\blauw{Fig.~\ref{fig5}d}) relative to the shorter (\blauw{Fig.~\ref{fig3}d}), and is the origin of both the increased breadth of the probability distribution and its long tail. The 50-mer catenane system possesses an interesting bimodal probability distribution with peaks at COM separations of 0.15 nm and 0.35 nm. To resolve the origin of these two peaks, we present in \blauw{Fig.~\ref{fig7}b} the projection of the 50-mer catenane ring configurations on the P-dMaps intrinsic manifold colored by the COM separation for the system. This plot reveals that the upper peak is associated with configurations in which one of the rings in the supramolecular complex is folded into a collapsed twisted boat configuration that causes the interlocked rings to tend to separate towards COM $\sim$0.35 nm. Conversely, when one of the rings is in a non-helical open or regular boat configuration, then the rings tend to be pulled together to COM $\sim$0.15 nm.

\subsection*{3.4 Deformability and relaxation rates}

To characterize the dynamical properties of each system, we computed for each simulation trajectory the time-averaged RMSD between united atom coordinates of the system as a function of the delay time $\tau$, $\langle \text{RMSD}(\textbf{r}(t),\textbf{r}(t+\tau)) \rangle$, where $\textbf{r}(t)$ is the configuration of the ring system at time $t$, RMSD($\textbf{r}_1$,$\textbf{r}_2$) measures the translational and rotational optimized root mean squared distance between configuration $\textbf{r}_1$ and $\textbf{r}_2$. We compute this quantity for the 24-mer and 50-mer chains and calculate it for the complete supramolecular system (i.e., linear chain, isolated ring, trefoil knot (averaged over the left- and right-handed chiralities), 2-ring catenane assembly, and 3-ring Borromean ring assembly).

\begin{figure*}[ht!]
\includegraphics[width=0.45\textwidth]{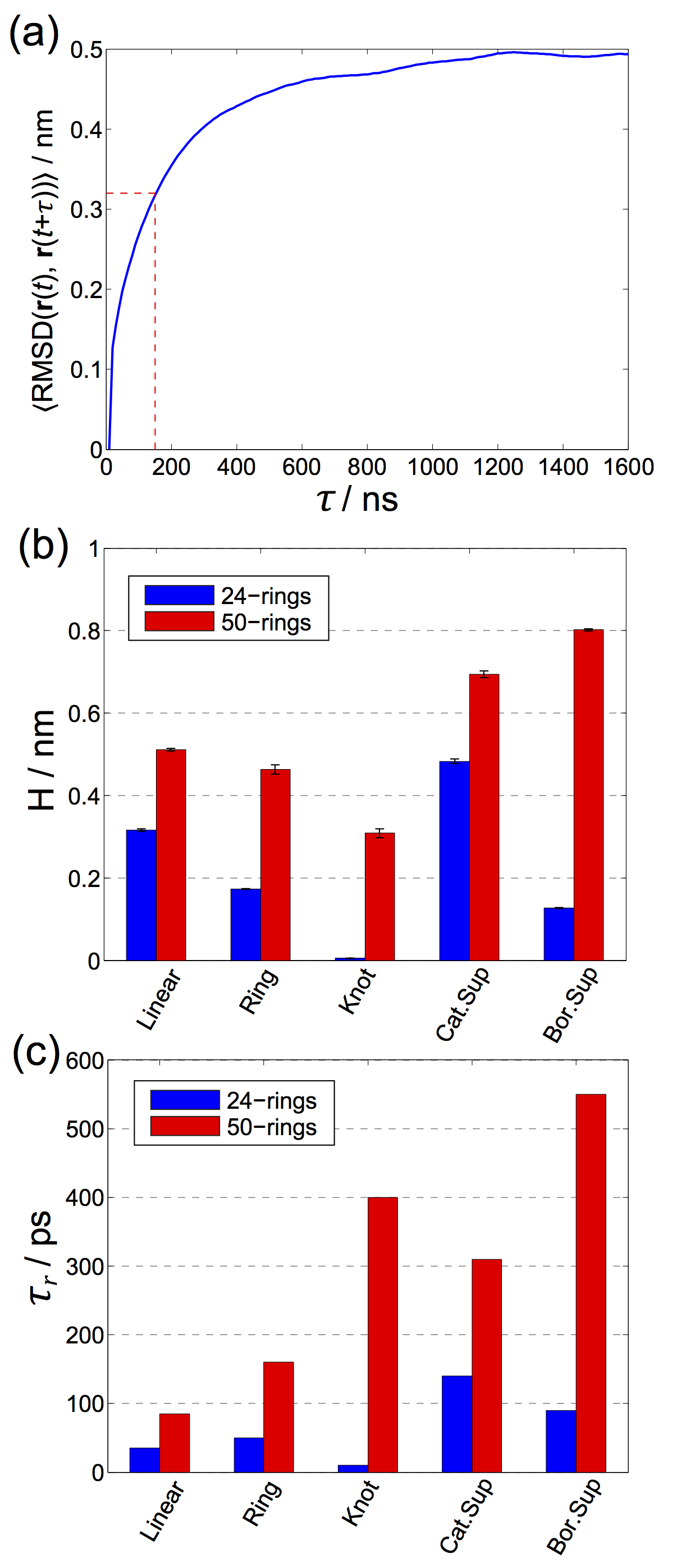}
\caption{\label{fig8} Deformability and relaxation rates as a function of supramolecular topology. (a) Plot of the time-averaged RMSD between time-delayed system configurations $\langle \text{RMSD}(\textbf{r}(t),\textbf{r}(t+\tau)) \rangle$ as a function of the delay time $\tau$ for the 50-mer isolated ring. The height of the plateau at $\tau \rightarrow \infty$ is termed $H$ and provides a measure of the average system deformability. The delay time at which $(1-\frac{1}{e})H$ is attained is defined as the relaxation time $\tau_r$ measures the timescale on which the system loses memory of its initial configuration and is illustrated by the dashed red line. (b) Plateau heights $H$ and (c) relaxation times $\tau_r$ for the linear, isolated ring, trefoil knot, 2-ring catenane assembly, and 3-ring Borromean ring assembly.}
\end{figure*}

We present in \blauw{Fig.~\ref{fig8}a} $\langle \text{RMSD}(\textbf{r}(t),\textbf{r}(t+\tau)) \rangle$ vs.\ $\tau$ for the 50-mer isolated ring. All other systems have the similar trends. $\langle \text{RMSD}(\textbf{r}(t),\textbf{r}(t+\tau)) \rangle$ increases from zero with increasing $\tau$ as the system loses memory of its original configuration and structurally decorrelates. To approach a plateau of height $H$ as $\tau \rightarrow \infty$. $H$ measures the average RMSD between uncorrelated system configurations, and may be taken as a measure of the deformability or flexibility of the system \cite{KUZMANIC2010}. Large $H$ values are indicative of large molecular flexibility and a large configurational space for molecular deformations, whereas small $H$ values correspond to a configurational ensemble of limited structural diversity. We define the delay time at which $\langle \text{RMSD}(\textbf{r}(t),\textbf{r}(t+\tau)) \rangle$ attains a height of $(1-\frac{1}{e})H$ as the relaxation time $\tau_r$. Large $\tau_r$ values indicates slow relaxation dynamics, whereas small values mean that the system rapidly forgets its initial structure. We present in \blauw{Fig.~\ref{fig8}b,c} the calculated values of $H$ and $\tau_r$ for the various supramolecular topologies.

The plateau heights $H$ for each topology are larger for the 50-mer chains compared to the 24-mers as expected from the greater configurational diversity accessible to the longer chains. For both chain lengths, the ordering of $H$ values among the single chains is trefoil knot $<$ isolated ring $<$ linear chain, due to the successive relaxation of topological constraints that limit the configurational ensembles accessible to the ring and knot relative to the linear chain (\blauw{Fig.~\ref{fig2}}, \blauw{Fig.~\ref{fig4}}). For the 50-mer chains, the catenane and Borromean ring supramolecular complexes have larger $H$ values than any of the single chain systems, indicative of the larger structural deformations that these large complexes can sustain. For the 24-mer chain, the 2-ring catenane complex has a larger $H$ value that reflects the larger configurational ensemble accessible to the two interlocked rings, but the $H$ value for the 3-ring Borromean ring complex lies below that of the isolated ring but above that of the trefoil knot. This observation reflects the very rigid and inflexible complex resulting from the assembly of such short chains into a Borromean ring topology.

The relaxation times for the single chain 24-mer polymers in linear, ring, and trefoil knot topologies all lie at $\tau_r$ $<$ 50 ps, indicating a very rapid structural relaxation. The relaxation times for the catenane and Borromean ring complexes of 24-mer chains is only slightly larger at $\tau_r$ $\approx$ 100 ps. For the 50-mer chains, the linear and isolated ring topologies also possess short relaxation times of 85 ps and 150 ps, respectively. Conversely, the 50-mer trefoil knot, 2-ring catenane complex, and 3-ring Borromean ring complex possess long relaxation times of 400 ns, 300 ns, and 550 ns, respectively. Accordingly, long relaxation times and long memory effects are favored by large chains that interact with one another in multi-chain complexes, or with themselves in highly constrained single chain topologies.

\subsection*{3.5 Rotational diffusivity}

\begin{figure*}[ht!]
\includegraphics[width=0.55\textwidth]{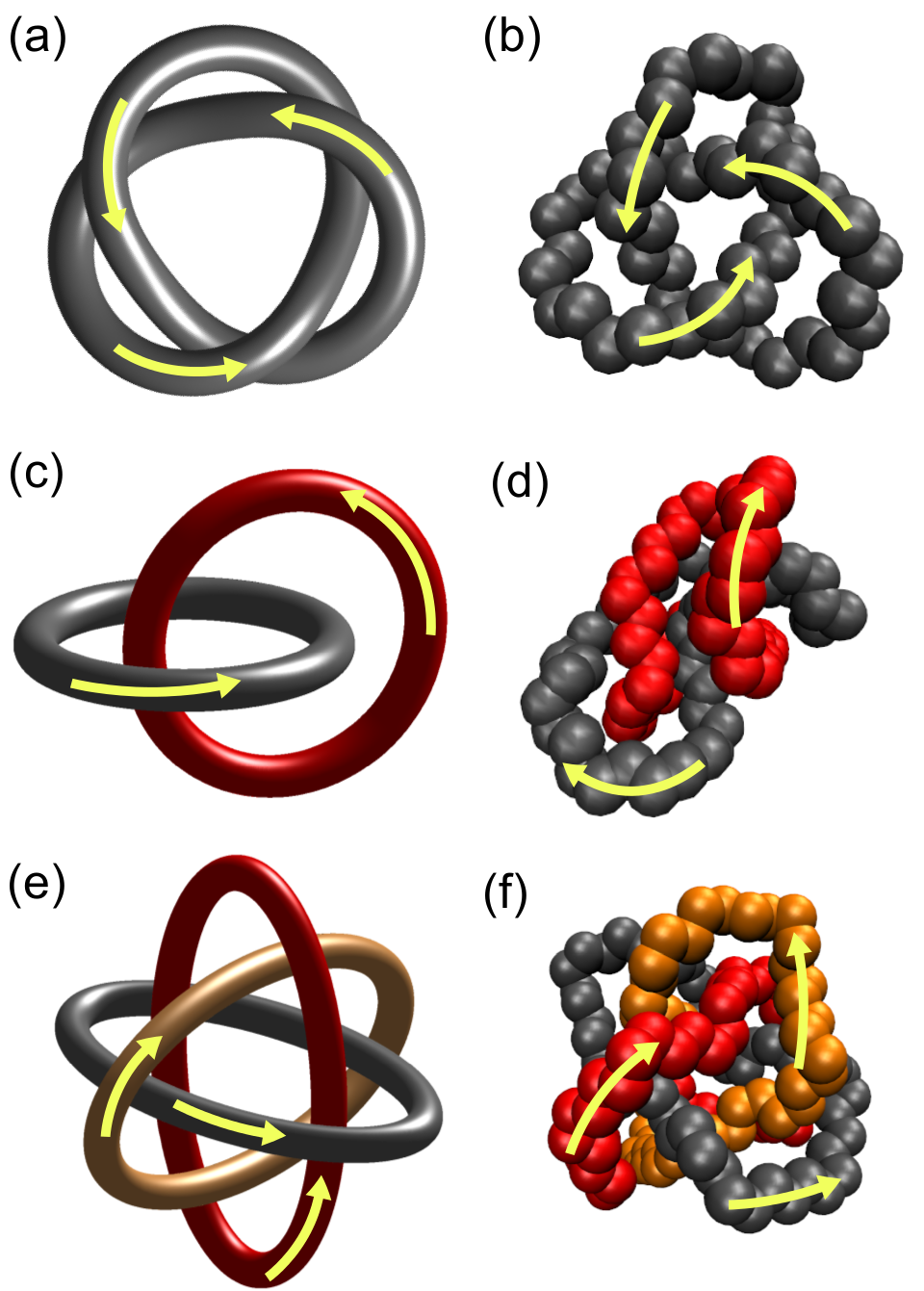}
\caption{\label{fig9} Illustration of intramolecular rotation within supramolecular topologies. Schematic and molecular illustrations of relative rotations or 50-mer polymer chains within (a,b) a trefoil knot, (c,d) catenane, and (e,f) Borromean ring.}
\end{figure*}

The constituent chains within supramolecular assemblies can rotate relative to one another within the complex such that -- under pure rotational motion -- the overall supramolecular structure remains unchanged but the relative position between the chains is altered. These relative motions, and their coupling to global motions of the supramolecular assembly, can play an important role in the behavior of the assembly. In assemblies composed of heteropolymers or chains containing functional groups, different internal rotational states may possess distinct structures or functions. For example, if the chains are engineered to hold photo-, electro-, or chemical-switchable groups, then the complex may be actuated and/or fixed in particular rotational conformations by applied external stimuli, and the complex may serve as an activated nano switch \cite{Switchlight, slidecatenane, switchelectro}. Even in the absence of external stimuli or fields, thermal fluctuations drive diffusive motions that alter the relative rotational state of the complex. We describe in this section a means to quantify this diffusive behavior and determine the impact of supramolecular topology on these dynamics. We present results for each of the two chain lengths in each of the three topologies for which relative rotations are possible: trefoil knot, catenane, and Borromean ring \blauw{Fig.~\ref{fig9}}.

To quantify the diffusive behavior, we need a means to distinguish the relative rotational states of the chains within the assembly between successive snapshots in the molecular simulation trajectory. To do so, we compute using the Kabsch algorithm \cite{kabsch1976solution} the RMSD between two successive observations of the $n$-chain supramolecular complex minimized over spatial translation, spatial rotation, and indexing of the $N$ monomers in each constituent chain, where $n=1$ for trefoil knot, 2 for catenane and 3 for Borromean system, $N$ is either 24 or 50. From this we can identify the optimal permutational indexing $0 < \Gamma_i < (N-1)$ of each chain in the complex. In principle, all $N^n$ indexings should be exhaustively searched in the optimization. In practice, since we are considering successive frames in the simulation trajectory separated by $\Delta t$ = 10 ps, the indexing rotation is expected to be small, so we can perform a fast search over only those local permutations. Empirically, we find searching up to $\pm 8$ index shifts is sufficient to robustly identify the global optimum. 

It is possible that large structural changes in the chains constituting the complex could mask our ability to distinguish relative rotational motions within our RMSD minimization. However, the relaxation times of the 24-mer catenane and Borromean ring, and 50-mer trefoil knot, catenane, and Borromean ring, are all in excess of $\sim 100$ ps (\blauw{Fig.~\ref{fig8}c}), which is much longer than the 10 ps delay between successive observations in the molecular simulation trajectory. Accordingly, the time scale of supramolecular structural relaxation is far longer than the separation between successive observations, and we can expect the conformational flexibility of the chains within complex will not significantly perturb our ability to distinguish their rotational states. The only exception to this separation of time scales is the 24-mer trefoil knot, which possesses a relaxation time of $\tau_r$  $\approx$ 10 ps = $\Delta t$ . However, we have shown this knot topology to be extremely rigid with $H$ = 0.005 nm (\blauw{Fig.~\ref{fig8}b}) such that the magnitude of conformational fluctuations is expected to be so small as not to interfere with the RMSD calculation. As we shall see, the rigidity of this topology is such that it prohibits any rotational motions and its rotational diffusivity is effectively zero.

Having defined the rotational state of each chain within the supramolecular complex in each frame of the molecular simulation trajectory, we model its dynamics as a discrete one-dimensional random walk with periodic boundary conditions \cite{Lemons, randomwalk}. The master equation describing the random walk is,
\begin{align}
P(t,x) &= p_0(\Delta t) P(t-\Delta t,x) + \sum_{k=1}^{\infty} p_k(\Delta t) P(t-\Delta t,x-kl) + \sum_{k=1}^{\infty} p_{-k}(\Delta t) P(t-\Delta t,x+kl) \notag \\ &= \sum_{k=-\infty}^{\infty} p_k(\Delta t) P(t-\Delta t,x-kl), \label{eqn:master}
\end{align}
where $P(t,x)$ is the probability that the walker is located at rotational state $x \in [0,(N-1)]$ at time $t$, $p_0(\Delta t)$ is the probability that the walker remains in its current position after the step time $\Delta t$, $p_k(\Delta t)$ is the probability it hops $k$ steps of size $l$ to the right, $p_{-k}(\Delta t)$ that it hops $k$ steps of size $l$ to the left, and $\sum_{k=-\infty}^{\infty} p_{k}(\Delta t) = 1$. We implicitly enforce periodicity such that the walker remains on the interval $[0,(N-1)]$. This expression explicitly allows for the possibility of any length of step within the time interval $\Delta t$. This is important for properly modeling our discrete random walk with step size $l$ is set by the separation between monomers. Specifically, depending on the observation interval $\Delta t$, we may expect to see 0, 1, 2, 3, ... steps of the random walker. 

As demonstrated in the \blauw{Supplementary Information}, for a symmetric random walk (i.e., $p_k = p_{-k}, \; \forall k$) in the limit $l \rightarrow 0$ and $\Delta t \rightarrow 0$, we identify the diffusion coefficient associated with \blauw{Eqn.~\ref{eqn:master}} as
\begin{equation}
D = \sum_{k=1}^{\infty}\frac{(kl)^2}{\Delta t}p_k. \label{eqn:D}
\end{equation}
We solve for $D$ by numerically extracting the $\{p_k\}$ from our calculated $\Gamma_i$ values over the simulation trajectory. To do so we compile histograms of jump sizes within the time interval $\Delta t$, symmetrize the $+k$ and $-k$ counts to enforce no preferred rotational directionality, and normalizing the distribution. We observe that in the limit that $\Delta t \rightarrow 0$ and $l \rightarrow 0$ only single jumps are permitted in each direction with equal probability (i.e., $p_1$ = $p_{-1}$ = 1/2, $p_{k \neq \{1,-1\}}$ = 0) and \blauw{Eqn.~\ref{eqn:D}} reduces to the more familiar expression \cite{Lemons},
\begin{equation}
D = \frac{l^2}{2\Delta t}. \label{eqn:Dapprox}
\end{equation}

The spatially discrete nature of our system means that we cannot take $l \rightarrow 0$ in tandem with $\Delta t \rightarrow 0$ to achieve a well defined limit for $D$ in \blauw{Eqn.~\ref{eqn:D}}. Specifically, the numerical calculation will break down for sufficiently small $\Delta t$ where the time interval too short for the fixed step size $l$ and $p_k = 0$, $\forall k$. Accordingly, we instead compute $D(\Delta t)$ for successively smaller values of $\Delta t$ and extrapolate to $\Delta t = 0$ using those points for which the approximation holds and the trend is smooth. We illustrate this procedure for the 50-mer chain catenane in \blauw{Fig.~S13}.

\begin{figure*}[ht!]
\includegraphics[width=0.70\textwidth]{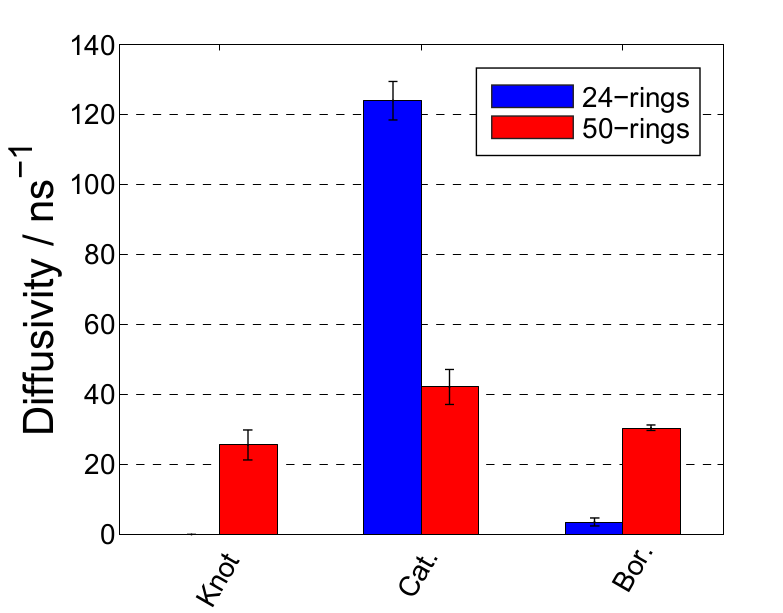}
\caption{\label{fig10} Diffusion coefficients for the 24-mer and 50-mer chains in the trefoil knot, catenane, and Borromean ring topologies. Values are computed using the $\Delta t \rightarrow 0$ extrapolation of \blauw{Eqn.~\ref{eqn:D}} as detailed in the main text. Values are reported in units of monomer rotations squared per ns. Error bars correspond to the standard error of the mean of five evenly blocked trajectories. The strong constraints imposed by the trefoil knot and Borromean ring topologies upon the 24-mer chains result in very small diffusivities of $D < 5$ ns$^{-1}$. The 50-mer chains in all three topologies possess intermediate rotational diffusivities of $D \approx 30$ ns$^{-1}$, and that of the 24-mer catenane is four times larger at $D = 125$ ns$^{-1}$.}
\end{figure*}

The rotational diffusion coefficient for the 24-mer and 50-mer chains in the trefoil knot, catenane, and Borromean ring topologies are presented in \blauw{Fig.~\ref{fig10}}. We adopt $l$ = 1 monomer, such that we measure distance in units of monomers and report diffusion constants in units of monomer rotations squared per unit time. For the 24-mer chains, the strong geometric constraints induced by the trefoil knot and Borromean ring topologies result in a very low rotational diffusivity of $D < 5$ ns$^{-1}$. This is consistent with the highly strained and unfavorable intra- and inter-molecular dispersion interactions reported in \blauw{Table~\ref{tb1}}. Conversely, the 24-mer chains in the catenane topology possess the highest rotational diffusivity of $D$ = 125 ns$^{-1}$. Employing the scaling relation for 1D diffusion $l^2 \sim 2Dt$ (\blauw{Eqn.~\ref{eqn:Dapprox}}), we can expect the rings within catenane to rotate by $\sim$15 monomer positions in a 1 ns time period. This is more than half the length of the ring, and indicates that the catenane topology admits very easy rotational motions. The structural basis for this is that there are no strong topological constraints or energetic interactions between the two interlocked rings (\blauw{Table~\ref{tb1}}), and the constituent rings tend to maintain very open nearly circular structures that readily admit free rotation (\blauw{Fig.~\ref{fig3}}).

For the 50-mer chains, all three topologies possess rotational diffusivities in the range 20-40 ns$^{-1}$, meaning that they are expected to rotate on the order of 6-9 monomer positions in 1 ns. The relatively larger mobilities of the 50-mer trefoil knot and Borromean ring relative to the 24-mer is due to the less stringent geometric and energetic constraints imposed by these topologies on the relatively longer chains (\blauw{Table~\ref{tb2}}). This preserves structural flexibility that permits the chains to sample a larger configurational ensemble and favors rotational motions. Interestingly, the 50-mer catenane rings possess a rotational diffusivity one third that of the 24-mers. This is a consequence of the larger intra-chain structural ensemble sampled by the larger ring that adopts collapsed configurations that disfavor free rotation due to structural entanglement of the two interlocked rings (cf.\ \blauw{Fig.~\ref{fig3}c}, \blauw{Fig.~\ref{fig5}c}).

\section*{4. Conclusions}

In this study, we performed molecular dynamics simulations of 24-mer and 50-mer polyethylene chains in water in a variety of supramolecular topologies: linear chain, isolated ring, right- and left-handed trefoil knot, catenane, and Borromean ring. We introduced an efficient variant of the diffusion map manifold learning technique termed pivot diffusion maps (\groen{P-dMaps}) that greatly reduces the computational and memory burden of the approach by efficient on-the-fly selection of a small number of pivot points with which to construct the low-dimensional embedding. Application of P-dMaps to the six supramolecular topologies of the 24-mer chain system discovered a 4D intrinsic manifold that revealed the relationships between the structural ensembles sampled in each of the topologies. The ring systems (isolated ring, catenane, Borromean ring) explored a subset of the configurations sampled by the linear chain, while the trefoil knots sampled a structurally distinct region of the manifold. For the 50-mer chains, P-dMaps identified an effectively 2D manifold that revealed the isolated ring configurations to be nested within that of the linear chain, and an overlapping relationship between linear, catenane, and Borromean ring. Again, the trefoil knots sampled a distinct region of the manifold that was inaccessible to the other topologies. The free energy surfaces constructed over the intrinsic manifold illuminated the relative stabilities of the chain configurations sampled in the various topologies. This analysis revealed a low-free energy pathway linking the circular and elongated ring configurations in the 24-mer isolated ring, along with slightly less stable twisted and boat configurations. For 50-mers, the collapsed states with right/left helicity are relatively stable, and are connected by free energy pathways containing achiral configurations. A decomposition of the potential energy experienced by the chain in the various topologies revealed extremely unfavorable intra- and inter-molecular dispersion interactions in the 24-mer trefoil knot and Borromean ring, respectively, imposed by the topology of the supramolecular assembly. The magnitude of these interactions suggest that these chemistries would be chemically unstable, but present a useful comparison for the 50-mer topologies where the twice as long chain is chemically stable and far less strained by the imposed topology.

An autocorrelation analysis of the chain RMSD enabled us to measure the deformability and relaxation rates of the chains in the various topologies. In general, the larger chains are both more deformable and exhibit slower relaxation rates, and demonstrated that longer memory effects are induced by increasing chain length or through inter-chain interactions in multi-chain complexes. We also proposed a means to compute the rotational diffusivity of the chains in trefoil knot, catenane, and Borromean ring complexes. Interestingly, while the 50-mer chain has a higher rotational mobility than the 24-mer in the knot and Borromean ring, the shorter chain possesses a three-fold larger rotational diffusivity in the catenane assembly. This is a consequence of the more open circular ring conformations sampled by the 24-mer compared to the collapsed structures explored by the 50-mer that impede rotation. 

In future work, we propose to conduct a more comprehensive study of the effect of degree of polymerization on chain structure and dynamics. The marked differences in the structural ensemble, energetic stability, and rotational diffusivity and relaxation time scales between the 24-mer and 50-mer chains suggests that the fundamental structure and dynamics of supramolecular complexes can be controllably tuned through the supramolecular topology and chain length. In particular, the unexpected higher rotational diffusivity of the 24-mer chain compared to the 50-mer in catenane suggests a non-monotonic dependence on degree of polymerization, and the existence of an optimal chain length to maximize rotational mobility. Quantitative resolution of this trend is of fundamental importance in the engineering molecular machines based on these chemistries such as nano-switches or molecular motors. We also propose that it would be of interest to deploy these techniques to the study of these complexes in non-aqueous solvents and in the melt state, in more complex supramolecular topologies such as large $n$-fold knots, Solomon links, and Whitehead links \cite{knotatlas}, and employing more complex polymer chemistries that are more representative of those used in the experimental fabrication of supramolecular systems \cite{Sauvage2007, balzani2000artificial, amabilino1995interlocked, collier20002, balzani1998molecular, pease2001switching, raymo1999interlocked}. In sum, this study provides new fundamental understanding of the impact of supramolecular topology on the fundamental structural and dynamic properties of polymer chains. We anticipate that this understanding, along with the computational analysis tools we introduced to derive it, will prove valuable in laying the foundational knowledge for the rational design and engineering of functional materials and devices based on mechanically constrained macromolecular chemistries.

\section*{Acknowledgments}

Acknowledgement is made to the Donors of the American Chemical Society Petroleum Research Fund (ACS PRF \# 54240-DNI6) for support of this research.



\clearpage
\newpage




\bibliography{Rings}

\clearpage
\newpage
\section{Graphical TOC Entry}
\begin{figure*}[ht!]
\includegraphics[width=0.85\textwidth]{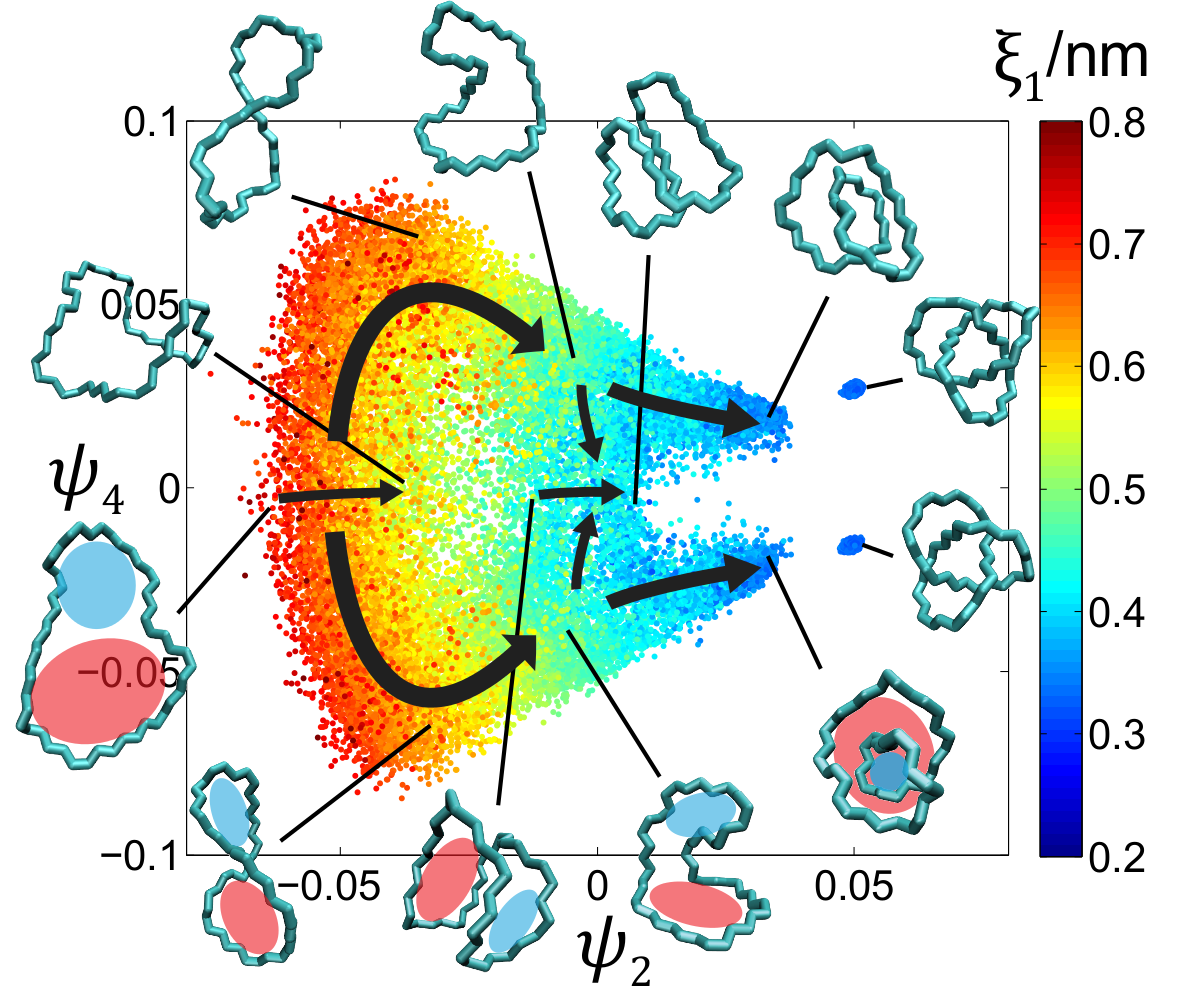}
\end{figure*}

\end{document}